\documentclass[showpacs,aps,prc,longbibliography,nofootinbib,showkeys,superscriptaddress,twocolumn]{revtex4-1}

\usepackage[colorlinks, urlcolor=blue,linkcolor=blue, anchorcolor=blue, citecolor=blue]{hyperref}
\usepackage[utf8]{inputenc}
\usepackage{natbib}
\usepackage{graphicx}
\usepackage{xcolor}
\usepackage{amsmath}
\usepackage{bm}
\usepackage{multirow}
\usepackage{array}

\newcommand{\music}{{\sc MUSIC}}
\newcommand{\isd}{{\sc iS3D}}
\newcommand{\iss}{{\sc iSS}}
\newcommand{\urqmd}{{\sc UrQMD}}
\newcommand{\snn}{\sqrt{s_\mathrm{NN}}}
\newcommand{\etas}{$\eta_s$}

\usepackage{amsthm}
\usepackage{mathtools}
\usepackage{physics}
\usepackage{xcolor}
\usepackage{graphicx}
\usepackage{adjustbox}
\usepackage{placeins}
\usepackage[T1]{fontenc}
\usepackage{lipsum}
\usepackage{csquotes}

\graphicspath{{figs/}}

\begin{document}

%%%%%%%%%%%%%%%%%%%%%%%%%%%%%%%%%%%%%%%%%%%%%%%%%%%%%%%%%%%%%%%%%%%%%%%%%%%%%%%%%%%%%%%

\title{Rapidity scan with multistage hydrodynamic and statistical thermal models}

\author{Lipei Du}
\affiliation{Department of Physics, McGill University, Montreal, Quebec H3A 2T8, Canada}
\author{Han Gao}
\affiliation{Department of Physics, McGill University, Montreal, Quebec H3A 2T8, Canada}
\author{Sangyong Jeon}
\affiliation{Department of Physics, McGill University, Montreal, Quebec H3A 2T8, Canada}
\author{Charles Gale}
\affiliation{Department of Physics, McGill University, Montreal, Quebec H3A 2T8, Canada}

\date{\today}

\begin{abstract}
    We calibrate a (3+1)-dimensional multistage hybrid framework using the measured pseudo-rapidity distribution of charged particles and rapidity distribution of net protons for central Au+Au collisions at $\snn=7.7,\,19.6,\,62.4,\,200$ GeV. We then study the thermodynamic properties of the nuclear matter along the beam direction, and the phase diagram regions probed by the hadronization process near the chemical freeze-out. Using the rapidity-dependent thermal yields of identified particles with full rapidity coverage from the hybrid framework, we apply different scenarios of the statistical thermal model to extract the thermodynamic parameters at the freeze-out, with the known system properties from the hybrid model as a closure test. We find significant theoretical uncertainties in the thermal models when applied to regions away from midrapidity. We also propose a thermal model inspired by the hybrid approach that includes thermal smearing and longitudinal flow for the nuclear matter created at low beam energies.
\end{abstract}

%%%%%%%%%%%%%%%%%%%%%%%%%%%%%%%%%%%%%%%%%%%%%%%%%%%%%%%%%%%%%%%%%%%%%%%%%%%%%%%%%%%%%%%

\maketitle

%%%%%%%%%%%%%%%%%%%%%%%%%%%%%%%%%%%%%%%%%%%%%%%%%%%%%%%%%%%%%%%%%%%%%%%%%%%%%%%%%%%%%%%
\section{Introduction}
%%%%%%%%%%%%%%%%%%%%%%%%%%%%%%%%%%%%%%%%%%%%%%%%%%%%%%%%%%%%%%%%%%%%%%%%%%%%%%%%%%%%%%%

Mapping the phase diagram of Quantum Chromodynamics (QCD) using heavy-ion collisions is one of the primary goals of nuclear physics \cite{Bzdak:2019pkr,An:2021wof}. This includes finding and studying a critical point. From high to low beam energies, the baryon charge increases in the nuclear matter created during the collision, and the collisions can then be used to probe the higher baryon chemical potential regions of the phase diagram \cite{Stephanov:1998dy,Stephanov:1999zu}. Such an energy scan is achieved by carrying out heavy-ion collisions at different experimental facilities, such as the Large Hadron Collider (LHC) at CERN and the Relativistic Heavy-Ion Collider (RHIC) at Brookhaven National Laboratory. At RHIC, the devoted Beam Energy Scan (BES) program has concluded, and large data sets have been accumulated, and new analyses with high statistics are ongoing. Additional experiments at even lower beam energies are planned at the newly constructed FAIR and NICA facilities.

Because of the dynamical nature of the QCD fireballs created in heavy-ion collisions, careful model-to-data comparisons are necessary before affirmative conclusions can be made on the existence of the critical point \cite{Bzdak:2019pkr,An:2021wof}. Theoretical modeling generally involves multistage hybrid approaches describing the various phases of the collision, and remarkable success has been achieved in describing ultra-relativistic heavy-ion collisions at LHC and top RHIC energies \cite{JETSCAPE:2020shq,JETSCAPE:2020mzn}. At those energies, the fireballs have negligible baryon charge and boost invariance is a good approximation near midrapidity, so (2+1)-dimensional descriptions at zero baryon chemical potential are generally implemented. This is not the case for collisions at BES and even lower energies, where an intrinsically (3+1)-dimensional nuclear interpenetration stage occurs during which energy and baryon number get deposited in the fireball dynamically. This creates pressure gradients due to inhomogeneity in the beam direction, which drive longitudinal flow and break the Bjorken expansion \cite{Bjorken:1982qr}. As a result, the boost invariant approximation can no longer be considered as a suitable approximation \cite{Shen:2017bsr,Li:2018fow,Denicol:2018wdp,Du:2021zqz,Du:2021uxo,Shen:2022oyg}. Recently, (3+1)-dimensional multistage models have been developed to account for longitudinal bulk dynamics and have shown notable success in explaining rapidity-dependent observables \cite{Shen:2017bsr,Denicol:2018wdp,Shen:2022oyg,Shen:2020jwv,Du:2022yok}.

Statistical thermal models \cite{Braun-Munzinger:2003pwq,Tawfik:2014eba} are another type of model that can extract the thermodynamic properties of the QCD matter from the final particle yields without the need for dynamical simulations. These models usually assume that the heavy-ion collision creates a hadron resonance gas (HRG) at equilibrium; thus, the hadron yields resemble statistical equilibrium ensembles \cite{Becattini:2003wp,Tawfik:2004sw,Andronic:2008gu,Cleymans:2005xv,Andronic:2017pug}. The success of these models has been demonstrated at LHC energies, where they have been used to extract the chemical freeze-out line from identified particle yields at midrapidity, measured across different beam energies \cite{Braun-Munzinger:2003pwq,Tawfik:2014eba,STAR:2017sal}. There have also been efforts to apply these models to fit the particle yields away from midrapidity \cite{Biedron:2006vf,Broniowski:2007mu,Becattini:2007qr,Becattini:2007ci,Stiles:2006sa,Begun:2018efg} and to extract rapidity-dependent thermodynamic parameters.

Due to the varying properties of the nuclear medium along the beam direction in low-energy collisions, the rapidity scan method \cite{Karpenko:2018xam, Brewer:2018abr, Begun:2018efg, Du:2021zqz, Du:2021uxo}, which employs observables in different rapidity windows to explore the QCD phase diagram, has gained increasing attention in recent years. However, while the idea of a rapidity scan is compelling, the challenge arises from boost-non-invariant longitudinal flow and baryon charge transport \cite{Denicol:2018wdp, Li:2018fow, Du:2018mpf, Du:2021zqz, Shen:2020jwv, Shen:2022oyg, Du:2021uxo, Du:2022yok}. Even though it is known that boost invariance becomes a less suitable approximation at lower beam energies, many studies using the thermal models rely on the Bjorken flow \cite{Schnedermann:1993ws,E895:2001zms,E-0895:2003oas,Netrakanti:2005iy}. Thus, quantitatively constraining the longitudinal flow and identifying the rapidity windows where boost invariance remains a valid approximation in a realistic setup are crucial for the successful application of thermal models in rapidity scans. Moreover, the lack of quantitative constraints on both the initial distribution of baryon charge and its transport along rapidity adds complexity to the challenge, although some progress has been made recently \cite{Shen:2017bsr,Denicol:2018wdp,Li:2018fow,Du:2019obx,Du:2021zqz,Du:2021uxo,Shen:2020jwv,Du:2022yok,Schafer:2021csj,Shen:2022oyg,Cimerman:2023hjw}.

In this work, we perform calibration of a (3+1)-dimensional hybrid framework consisting of \music+\isd+\urqmd{} by using the pseudo-rapidity distribution of charged particles and the rapidity distribution of net protons, for the central Au+Au collisions at $\snn=7.7,\,19.6,\,62.4,\,200$ GeV, which span a wide range of the BES energies. With this framework, we investigate the thermodynamic properties of the fireball on the freeze-out surface along the beam direction and the regions in the QCD phase diagram explored by the hadronization process within the multistage hybrid approach. Especially, we also present the longitudinal flow which is faster than the Bjorken flow away from midrapidity because of the longitudinal pressure gradients.

Using the rapidity-dependent identified particle yields given by the hybrid model, with rapidity coverage not available in experiments, we apply three different statistical thermal model scenarios to extract the rapidity-dependent thermodynamic parameters of the fireball at the chemical freeze-out. Additionally, we examine the effects of the rapidity spread due to the random thermal motion of the radiated particles and the boost-non-invariant longitudinal flow on the particle distributions. Drawing inspiration from the hybrid approach, we propose a thermal model that accounts for both of these effects. By utilizing the freeze-out surface obtained from the hybrid approach, we evaluate the performance of different thermal model scenarios.

This paper is organized as follows. In Sec.~\ref{sec:models}, we describe the settings of the multistage hybrid framework and different scenarios of the thermal model used in this study. Then we present and compare the results of the two approaches in Sec.~\ref{sec:results}, including the rapidity dependence of the thermodynamic properties of the fireball at freeze-out and their distributions in the QCD phase diagram. The concluding remarks are given in Sec.~\ref{sec:conclusions}. 

%%%%%%%%%%%%%%%%%%%%%%%%%%%%%%%%%%%%%%%%%%%%%%%%%%%%%%%%%%%%%%%%%%%%%%%%%%%%%%%%%%%%%%%
\section{Models and setup}\label{sec:models}
%%%%%%%%%%%%%%%%%%%%%%%%%%%%%%%%%%%%%%%%%%%%%%%%%%%%%%%%%%%%%%%%%%%%%%%%%%%%%%%%%%%%%%%

%%%%%%%%%%%%%%%%%%%%%%%%%%%%%%%%%%%%%%%%%%%%%%%%%%%%%%%%%%%%%%%%%%%%%%%%%%%%%%%
\subsection{Multistage hydrodynamic model}\label{sec:hybrid_setup}
%%%%%%%%%%%%%%%%%%%%%%%%%%%%%%%%%%%%%%%%%%%%%%%%%%%%%%%%%%%%%%%%%%%%%%%%%%%%%%%

A (3+1)-dimensional hybrid framework with parametric initial conditions is used to simulate central Au+Au collisions at BES energies, following Refs.~\cite{Denicol:2018wdp,Du:2022yok}. We construct the initial conditions of entropy and baryon densities by extending the nucleus thickness function with parametrized longitudinal profiles in a manner similar to that described in Ref.~\cite{Denicol:2018wdp}. However, unlike Ref.~\cite{Denicol:2018wdp} which assumes that the net baryon charge deposited into the system is equal to the total number of participant nucleons $N_\mathrm{part}$, we introduce an extra factor $N_B$ and assume the net baryon charge is $N_BN_\mathrm{part}$. The reason is that using $N_\mathrm{part}$ as the net baryon charge can strongly overestimate the net proton yields away from midrapidity at 200 GeV \cite{Shen:2020jwv}. The plateau component introduced in the initial baryon profile by Ref.~\cite{Du:2022yok}, which is found to be essential for explaining the rapidity-dependent directed flows of baryons, is not included in this study, as we are mainly interested in the distributions of temperature and baryon chemical potential at the freeze-out which are not sensitive to such a component.

The hydrodynamic stage is initialized at a constant proper time $\tau_0$ with Bjorken flow, and thus the hydrodynamic flow is driven by pressure gradients. The energy-momentum tensor and net baryon current are propagated with the dissipative effects from the shear stress tensor and net baryon diffusion current \cite{Denicol:2018wdp,Du:2022yok}, simulated by \music{} \cite{Schenke:2010nt,Schenke:2011bn,Paquet:2015lta}. We use a specific shear viscosity $\eta/s$ which has both temperature $T$ and baryon chemical potential $\mu$ dependence \cite{Shen:2020jwv}, and a baryon diffusion coefficient $\kappa$ which is obtained from the Boltzmann equation in the relaxation time approximation in the massless limit \cite{Denicol:2018wdp}
\begin{equation}
    \kappa = \frac{C_B}{T} n \left(\frac{1}{3} \coth\left(\frac{\mu}{T}\right) - \frac{n T}{e + P}\right)\,,
\end{equation}
where $e, n, P$ are energy density, baryon density, and pressure, respectively. In this study, we set $C_B=0.3$ \cite{Denicol:2018wdp,Du:2021zqz}, while Ref.~\cite{Du:2022yok} uses a smaller value $C_B=0.1$. We use an equation of state (EoS) at vanishing strangeness and electric charge chemical potentials, NEOS-B \cite{Monnai:2019hkn}, which smoothly interpolates between the result obtained using lattice QCD EoS at high $T$, and that associated with a hadron resonance gas one at low $T$. 

When the system expands and cools, the particlization process is carried out on a freeze-out hypersurface at a constant energy density $e_\mathrm{fo}{\,=\,}0.26$ GeV/fm$^3$ \cite{Shen:2020jwv}. We implement \isd{} \cite{McNelis:2019auj} to sample hadrons on the freeze-out surface using the Cooper-Frye prescription \cite{Cooper:1974mv}, including the off-equilibrium effects from shear stress and baryon diffusion. Performance of the  \isd{} sampler is validated by comparing to \iss{} \cite{Shen:2014vra} (see Fig.~\ref{fig:is3d_vs_iss} below). The hadronic afterburner is simulated by a kinetic transport description, \urqmd{} \cite{Bass:1998ca,Bleicher:1999xi}. Weak decay feed-down contributions are included in the yield of net protons when compared to the experimental data.

%%%%%%%%%%%%%%%%%%%%%%%% - tab - %%%%%%%%%%%%%%%%%%%%%%%%
\begin{table}[!btp]
\centering
\begin{tabular}{ccccccccc}
\hline
\hline
$\sqrt{s_\mathrm{NN}}\, \mathrm{(GeV)}$ & $\tau_0\, \mathrm{(fm)}$ & $s_0$ & $\eta^s_0$ & $\sigma_{\eta,s}$ & $N_B$ & $\eta^B_0$ & $\sigma_{\eta,+}$ & $\sigma_{\eta,-}$ \\
\hline
7.7 & 3.6  & 2.44 & 0.75& 0.17 & 1.22 & 1.28 & 0.07 & 0.75 \\
19.6 & 1.8 & 5.56 & 1.5 & 0.25 & 0.97 & 1.7 & 0.07 & 0.93 \\
62.4 & 1.0 & 12.1 & 2.3 & 0.3  & 0.93 & 2.8 & 0.21 & 1.3 \\
200 & 1.0  & 16.6 & 2.5 & 0.6  & 0.5  & 2.9 & 0.5 & 1.5 \\
\hline
\hline
\end{tabular}
\caption{Parameters of the initial conditions used by the hybrid model for the results shown in Figs.~\ref{fig:bes_yields_tune} and \ref{fig:is3d_vs_iss}. The same parametrization and symbols are used as in Ref.~\cite{Denicol:2018wdp}. See the text for more details.}
\label{tab:init}
\end{table}
%%%%%%%%%%%%%%%%%%%%%%%%%%%%%%%%%%%%%%%%%%%%%%%%%%%%%%%%%

%%%%%%%%%%%%%%%%%%%%%%%% - fig - %%%%%%%%%%%%%%%%%%%%%%%%
\begin{figure}[!tbp]
    \centering
    \includegraphics[width=\linewidth]{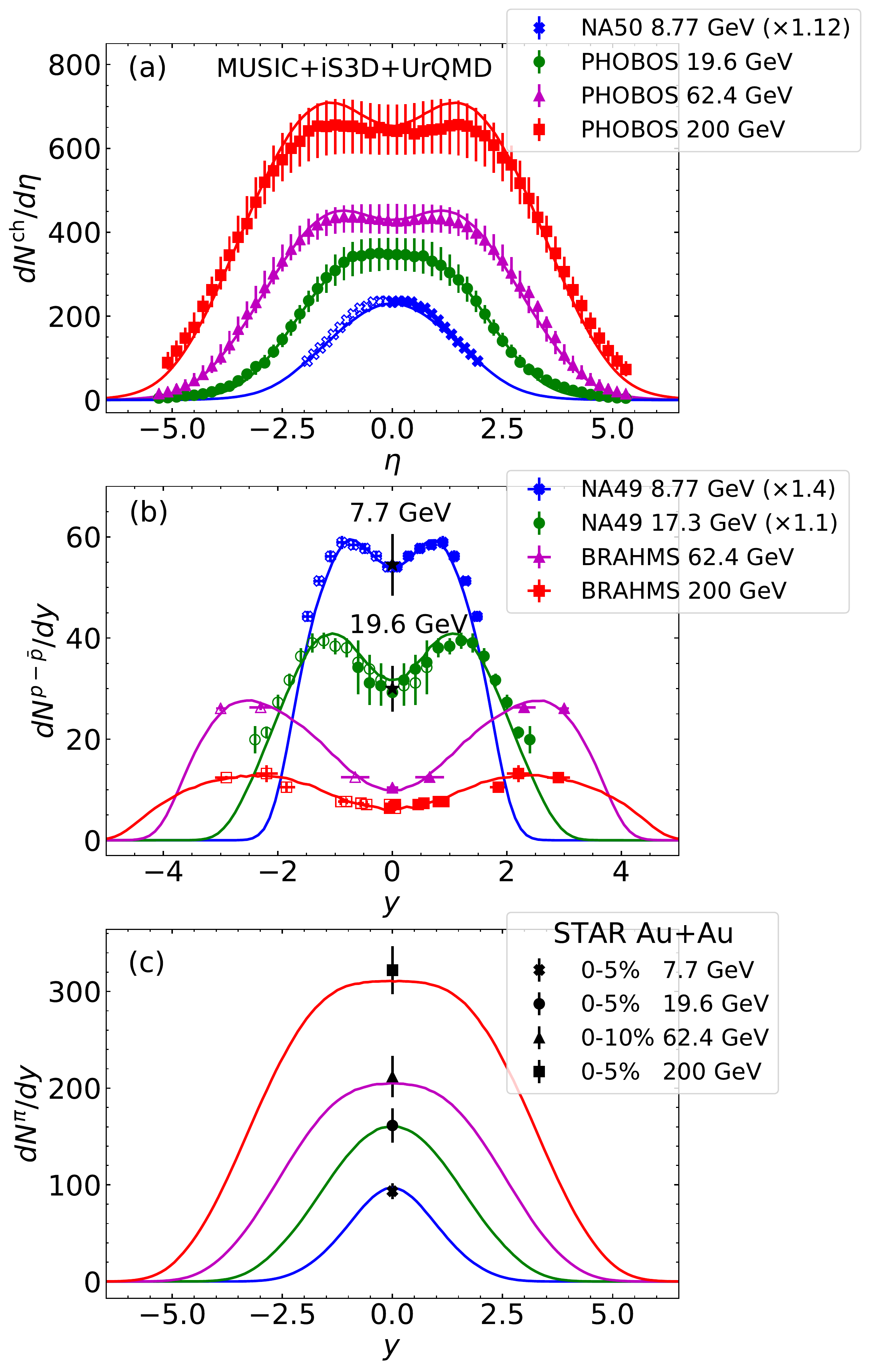}
    \caption{Comparison of the measured (a) pseudo-rapidity distribution of charged particle multiplicity, and rapidity distribution of (b) net protons and (c) pions for central Au+Au collisions at $\snn=7.7,\, 19.6,\, 62.4$ and 200 GeV to the theoretical results from the hybrid model consisting of \music+\isd+\urqmd{}. The data points are labeled by the markers and the theoretical results by the solid lines. References of the  experimental measurements are listed in Table \ref{tab:data}. See the text for more details.}
    \label{fig:bes_yields_tune}
\end{figure}
%%%%%%%%%%%%%%%%%%%%%%%%%%%%%%%%%%%%%%%%%%%%%%%%%%%%%%%%%

We use the pseudo-rapidity distribution of charged particles and net proton yields in rapidity to constrain the longitudinal bulk dynamics of central Au+Au collisions at $\snn=7.7,\,19.6,\,62.4,\,200$ GeV, in the same way as Ref.~\cite{Du:2022yok}. The tuned initial condition parameters are listed in Table~\ref{tab:init}. Comparison between the experimental measurements and results from the theoretical simulations are shown in Fig.~\ref{fig:bes_yields_tune}. The measured rapidity distribution of net protons at $\snn=200$ GeV only covers $|y|\lesssim3$, and thus the peaks in the complete distribution \cite{Mehtar-Tani:2008hpn,Alvarez-Muniz:2009qaj} near the beam rapidity $y_\mathrm{b}=5.36$ are not measured. We limit our fit to the available data points and thus miss the maximum peaks near the fragmentation regions. This would underestimate the maximum baryon chemical potential. In Fig.~\ref{fig:is3d_vs_iss}, we also calculate the invariant momentum spectra of identified particles at midrapidity for 0-5\% Au+Au collisions at 19.6 GeV, and find those to be in good agreement with the measurements. We also show the results obtained from the framework where \isd{} is replaced by \iss{} in Fig.~\ref{fig:is3d_vs_iss}, and excellent agreement is achieved, which validates the \isd{} sampler at the nonzero chemical potential.

%%%%%%%%%%%%%%%%%%%%%%%% - fig - %%%%%%%%%%%%%%%%%%%%%%%%
\begin{figure}[!btp]
\begin{center}
%\hspace{-.5cm}
\includegraphics[width=0.8\linewidth]{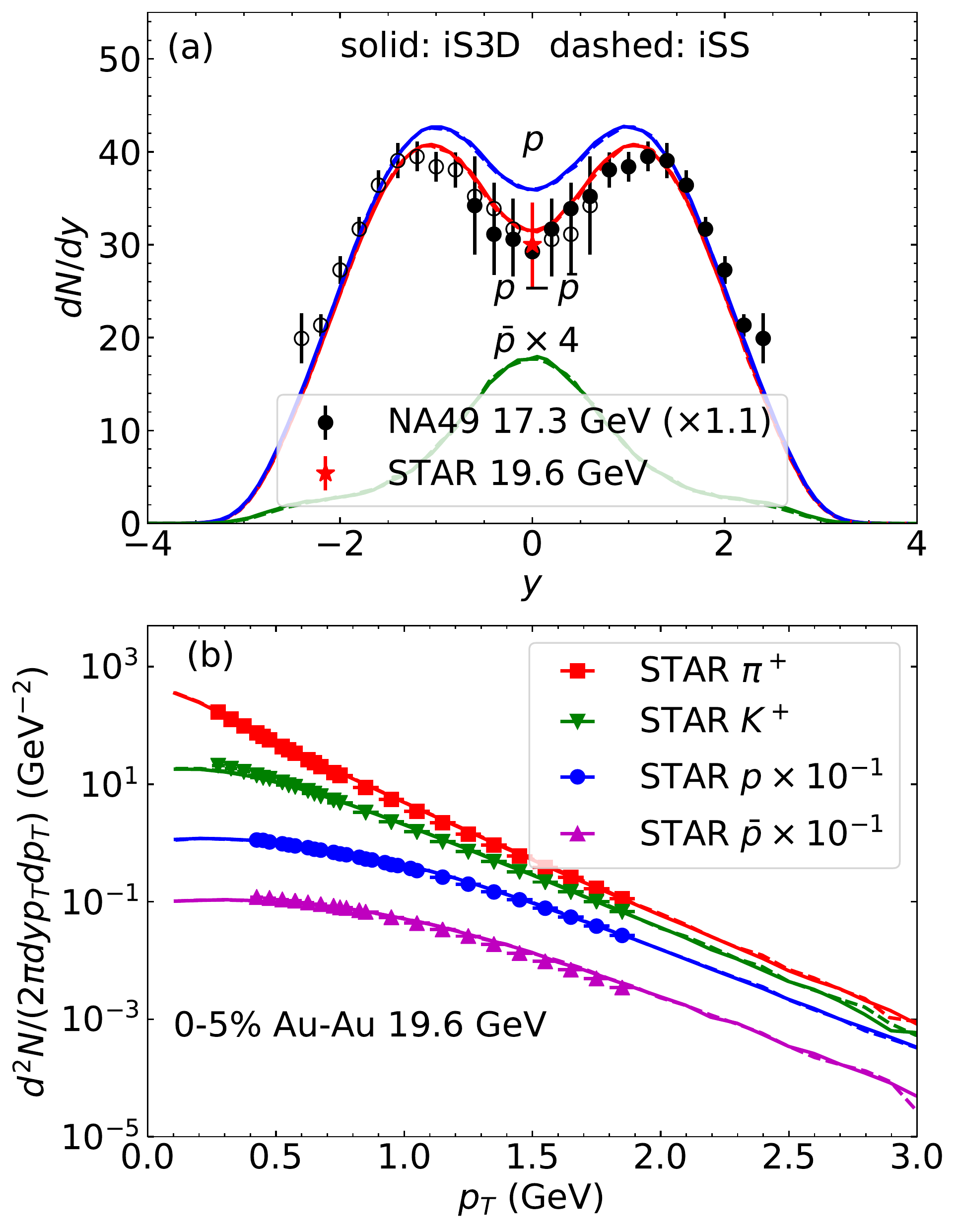}
    \caption{%
        (a) Rapidity distributions of protons $p$, anti-protons $\bar p$ and net protons $p-\bar p$, and (b) $p_T$ spectra of pions $\pi^+$, kaons $K^+$, protons $p$ and anti-protons $\bar p$ at midrapidity, for 0-5\% Au+Au collisions at $\snn=19.6$ GeV. The figure also shows the validation of the hybrid framework consisting of \music+\isd+\urqmd{} by comparing its results (solid) to those obtained by replacing \isd{} with \iss{} (dashed). Data points are from Refs.~\cite{NA49:1998gaz,STAR:2017sal}.  See the text for more details.
        }
    \label{fig:is3d_vs_iss}
\end{center}
\end{figure}
%%%%%%%%%%%%%%%%%%%%%%%%%%%%%%%%%%%%%%%%%%%%%%%%%%%%%%%%%

For the purpose of illustrating the connection between the multistage hydrodynamic and statistical thermal approaches below, we focus on the particlization process which converts hydrodynamic fields into hadronic resonances on the freeze-out surface $\Sigma$. One can compute the invariant momentum spectrum of each hadron species $h$ on $\Sigma$ using the Cooper-Frye prescription \cite{Cooper:1974mv}
\begin{equation}
    \label{eq:cooper-frye}
  E_p \frac{dN_h}{d^3p} = \frac{g_h}{(2\pi)^3}\int_\Sigma p \cdot d^3\sigma(x) \, f_h(x,p) \,,
\end{equation}
%
%
%%%%%%%%%%%%%%%%%%%%%%%% - table - %%%%%%%%%%%%%%%%%%%%%%%%
\begin{table*}[!tbp]
\centering
\begin{tabular}{cccc}
\hline
\hline
$\sqrt{s_\mathrm{NN}}$ & Collision system & Observable & Collaboration  \\ 
%\hline
    % 7.7
    \hline
    7.7 GeV & Au+Au & $dN/dy|_{|y|<0.1}$ of $\pi^+$, $K^+$, $p$ and $\bar{p}$ \cite{STAR:2017sal} & STAR\\
    
    % 8.77
    \hline
    \multirow{2}{*}{8.77 GeV}  & \multirow{2}{*}{Pb+Pb} & $dN/dy$ of $p$ and $\bar{p}$ \cite{NA49:2010lhg} and of $\pi^+$ and $K^+$ \cite{NA49:2012rsi}  & NA49 \\
    %\cline{3-4}
    & & $dN/d\eta$ of charged particles \cite{NA50:2002edr} & NA50\\
    % 17.3
    \hline
    17.3 GeV & Pb+Pb & $dN/dy$ of $p-\bar{p}$ \cite{NA49:1998gaz} and of $\pi^+$ and $K^+$  \cite{NA49:2012rsi} & NA49\\
    % 19.6 GeV
    \hline
    \multirow{2}{*}{19.6 GeV} & \multirow{2}{*}{Au+Au} & $dN/d\eta$ of charged particles \cite{Back:2002wb} &  PHOBOS\\ 
    %\cline{3-4}
    & & $dN/dy|_{|y|<0.1}$ and $p_T$ spectra of $\pi^+$, $K^+$, $p$ and $\bar{p}$ \cite{STAR:2017sal} & STAR\\
    % 62.4 GeV
    \hline
    \multirow{3}{*}{62.4 GeV} & \multirow{3}{*}{Au+Au} & $dN/d\eta$ of charged particles \cite{PHOBOS:2005zhy} &  PHOBOS\\ 
    %\cline{3-4}
    & & $dN/dy$ of $\pi^+$ and $K^+$ \cite{BRAHMS:2009acd}, and of $p$ and $\bar{p}$ \cite{BRAHMS:2009wlg}& BRAHMS\\
    %\cline{3-4}
    & & $dN/dy|_{|y|<0.1}$ of $\pi^+$, $K^+$, $p$ and $\bar{p}$ \cite{STAR:2008med} & STAR\\
    % 200 GeV
    \hline
    \multirow{4}{*}{200 GeV} & \multirow{4}{*}{Au+Au} & $dN/d\eta$ of charged particles \cite{Back:2002wb} &  PHOBOS\\ 
    %\cline{3-4}
    & & $dN/dy$ of $\pi^+$ and $K^+$ \cite{BRAHMS:2004dwr}, and of $p$ and $\bar{p}$ \cite{BRAHMS:2003wwg} & \multirow{2}{*}{BRAHMS}\\
    & & $dN/d\eta$ of charged particles \cite{BRAHMS:2001llo}& \\
    %\cline{3-4}
    & & $dN/dy|_{|y|<0.1}$ of $\pi^+$, $K^+$, $p$ and $\bar{p}$ \cite{STAR:2008med} & STAR\\
\hline
\hline
\end{tabular}
\caption{Measurements of identified particle yields and charged particle multiplicities for Au+Au and Pb+Pb collisions at various beam energies.}
\label{tab:data}
\end{table*}
%%%%%%%%%%%%%%%%%%%%%%%%%%%%%%%%%%%%%%%%%%%%%%%%%%%%%%%%%
%
%
where $g_h$ is the spin degeneracy factor and $f_h(x,p)$ the phase-space distribution function for the given hadron species, $E_p=\sqrt{m_h^2+p^2}$ is its energy with mass $m_h$ and four-momentum $p^\mu$, and $d^3\sigma_\mu(x)$ is the four-vector of a three-dimensional freeze-out hypersurface element with space-time coordinates $x$. In practice, we integrate the Cooper-Frye formula numerically \cite{McNelis:2019auj} on a discretized freeze-out surface, and the momentum spectra of hadrons emitted from a selected freeze-out cell $i$ are given as
\begin{equation}
\label{eq:continuous_spectra}
    E_p \frac{dN_h^i}{d^3p} = \frac{g_h}{(2\pi)^3} \bigl[p \cdot d^3\sigma(x_i)\bigr] \, 
    f_h(x_i, p)\,,
\end{equation}
where $d^3\sigma(x_i)$ is a discrete hypersurface element at position $x_i$. 

Decomposing particle's momentum into temporal and spatial components, $p^\mu=u^\mu(p\cdot u)+\Delta^{\mu\nu}p_\nu$, using the fluid velocity $u^\mu$ and spatial projector $\Delta^{\mu\nu}\equiv g^{\mu\nu}-u^\mu u^\nu$, we have $p \cdot d^3\sigma = (u\cdot d^3\sigma)(p\cdot u) + p\cdot \Delta\cdot d^3\sigma$. We can obtain the mean number of a hadron species emitted from the selected freeze-out cell by integrating out its momentum $p$:
\begin{align}
\label{eq:meanHadronsNoOutflow}
\Delta N_h(x) = &\bigl[u(x) \cdot d^3\sigma(x)\bigr] \left[g_h \int_p u(x) \cdot p \, f_h(x, p)\right]\nonumber\\
    &+ d^3\sigma(x) \cdot \left[g_h  \int_p \Delta\cdot p  \, f_h(x, p)\right]\,,
\end{align}
where $\int_p \equiv \int d^3p/[(2\pi)^3E_p]$. As expected, we see that the first term of Eq.~\eqref{eq:meanHadronsNoOutflow} is the product of the particle number density, $n_h(x)\equiv g_h \int_p u(x) \cdot p \, f_h(x,p)$, and the hypersurface volume element $V(x)\equiv u(x) \cdot d^3\sigma(x)$, in the local rest frame. The second term corresponds to the contribution of the diffusion current $n_h^\mu$ passing through the surface element, where $n_h^\mu(x)\equiv g_h \int_p \Delta^\mu_\nu p^\nu\, f_h(x,p)$.

At the hadronization of the multistage approach, by interpreting Eq.~\eqref{eq:continuous_spectra} as a probability density in phase space and sampling it stochastically for hadron resonances, the hydrodynamic fields $(T, \mu, u^\mu, \ldots)$ on the freeze-out surface $\Sigma$ are converted into particles with position and momentum. However, the final particles are only measured in momentum space, making it challenging to extract thermodynamic properties in coordinate space from these measurements, especially with the complications of longitudinal flow and thermal smearing, which we will discuss below.

%%%%%%%%%%%%%%%%%%%%%%%%%%%%%%%%%%%%%%%%%%%%%%%%%%%%%%%%%%%%%%%%%%%%%%%%%%%%%%%
\subsection{Statistical thermal model}\label{sec:thermal_setup}
%%%%%%%%%%%%%%%%%%%%%%%%%%%%%%%%%%%%%%%%%%%%%%%%%%%%%%%%%%%%%%%%%%%%%%%%%%%%%%%

To carry out the rapidity scan with the statistical thermal models, we would need to extract thermodynamic parameters along the beam direction from the rapidity-dependent measurements. This work considers three different thermal model scenarios: 
The first scenario treats the entire fireball as a thermal source, characterized by an effective temperature and baryon chemical potential. 
In the second scenario, particle species measured at rapidity $y$ are assumed to originate solely from an independent thermal source located at rapidity $y_s=y$.\footnote{%
    The subscript $s$ denotes that $y_s$ represents the rapidity associated with a thermal source, distinct from the rapidity of the particle denoted as $y$.
} 
The third scenario considers thermal smearing, and a thermal source at rapidity $y_s$ is capable of generating particles with a range of rapidities centered around $y=y_s$.

Treating a freeze-out cell in Eq.~\eqref{eq:continuous_spectra} as a thermal source, one can obtain the hadron yields with given thermodynamic parameters. If we ignore the off-equilibrium effects\footnote{%
    The off-equilibrium corrections are usually ignored in thermal models, and they can change the net proton yields by about 3\% for 0-5\% Au+Au collisions at $\snn=19.6$ GeV using the Cooper-Frye prescription on a hydrodynamic freeze-out surface \cite{Denicol:2018wdp}.
}
and thus the diffusion contribution in Eq.~\eqref{eq:meanHadronsNoOutflow} vanishes,\footnote{%
    The Landau matching conditions are commonly used to achieve this \cite{Jeon:2015dfa}.
}
then Eq.~\eqref{eq:continuous_spectra} becomes
\begin{equation}
\label{eq:singlecell}
    E_p \frac{dN_h}{d^3p} = \frac{g_h}{(2\pi)^3} \bigl[u(x) \cdot d^3\sigma(x)\bigr]\bigl[u(x) \cdot p\bigr] \, 
    f_h(x, p)\,.
\end{equation}
The invariant momentum spectrum of a particle species radiated by a static isotropic source can be written as:
\begin{equation}
\label{eq:singlecell_lrf}
E_p \frac{dN_h}{d^3p} = \frac{dN_h}{dy_hm_Tdm_Td\phi_h} =\frac{g_hV}{(2\pi)^3} E_pf_h\,,
\end{equation}
where $m_T\equiv\sqrt{m_h^2+p_T^2}$ is the transverse mass and $p_T,\phi_h, y_h$ are the transverse momentum, azimuthal angle, and rapidity of the hadron species, respectively, and $V$ the volume of the source at rest.  Integrating out the dependencies on transverse mass $m_T$ and azimuthal angle $\phi_h$ in Eq.~\eqref{eq:singlecell_lrf}, we can get the rapidity-dependent distributions of hadrons emitted by the source with $y_s=0$:\footnote{%
    From here, we suppress the subscript $h$ denoting hadrons in the hadron rapidity $y_h$.
}
\begin{equation}\label{eq:rap_dis}
    \left.\frac{dN_h}{dy}\right|_{y_s=0}=\frac{g_hV}{(2\pi)^3}\int_0^{2\pi}\int_{m_h}^\infty  E_pf_h(T, \mu)\,m_Tdm_Td\phi_h\,,
\end{equation}
where $(T, \mu, V)$ are thermodynamic properties of the source. Ignoring the off-equilibrium corrections, the single particle distribution function is
\begin{equation}\label{eq:1part}
    f_h(T, \mu)=\frac{1}{\exp\bigl[(E_p-\mu)/T\bigr]-\theta}\,,
\end{equation}
with $\theta=-1$ for fermions (Fermi-Dirac distribution) and $\theta=1$ for bosons (Bose-Einstein distribution).  

Since the proton mass is much larger than the freeze-out temperature, $m_p = 938{\rm \ MeV} \gg T \approx 160{\rm \ MeV}$, the $f_h$ in Eq.~\eqref{eq:1part} can be well approximated by the Maxwell-Boltzmann (MB) distribution with $\theta=0$, $f_p(T, \mu)=\exp\bigl(-(E_p-\mu)/T\bigr)$. Plugging $f_p(T, \mu)$ and $E_p=m_T\cosh y$  in Eq.~\eqref{eq:rap_dis} and carrying out the integrals, we obtain the rapidity distribution of protons \cite{Schnedermann:1993ws}:
\begin{align}\label{eq:p_yield}
    \left.\frac{dN_p}{dy}\right|_{y_s=0} = \frac{g_pVT^3}{(2\pi)^2} &\left( \frac{2}{\cosh^2 y} + \frac{m_p}{T}\frac{2}{\cosh y}
    + \frac{m^2_p}{T^2}\right)\nonumber\\
    &\times \exp\left(\frac{\mu-m_p\cosh y}{T}\right)\,,
\end{align}
where $g_p = 2$ is the spin degeneracy of the proton. Replacing $\mu$ by $-\mu$ in the above equation gives the anti-proton distribution. On the other hand, the masses of pions and kaons are of the same order as the freeze-out temperature, and thus the Bose-Einstein (BE) distribution is needed for the $f_h$ in Eq.~\eqref{eq:1part}. Expanding the BE distribution as a series of MB distributions, we can get the rapidity distributions of kaons and pions \cite{Denicol:2018wdp}:
\begin{align}\label{eq:pik_yield}
    \left.\frac{dN_i}{dy}\right|_{y_s=0} = &\frac{g_i VT^3}{(2\pi)^2}\sum_{n=1}^\infty\left(\frac{1}{n}\right)^3 \left( \frac{2}{\cosh^2 y} + \frac{nm_i}{T}\frac{2}{\cosh y} \right.\nonumber\\
    &\left.+ \frac{n^2m^2_i}{T^2}\right)\exp\left(-\frac{nm_i\cosh y}{T}\right)\,,
\end{align}
where $i=\pi^+, K^+$ with spin degeneracy factor $g_i = 1$. 

Eqs.~(\ref{eq:p_yield},\ref{eq:pik_yield}) indicate that a thermal source can emit particles with a range of rapidities due to particles' random thermal motion (i.e., thermal smearing effect) \cite{Schnedermann:1993ws}. For example, protons radiated from a thermal source have a Gaussian-like distribution in the rapidity with width $\Delta y \approx 0.9$, and pions and kaons have even larger rapidity spreads with $\Delta y \approx 1.6$ and 1.2, respectively \cite{Begun:2018efg}.

{\it \bf Single source model.} Some studies treat the entire fireball as a single thermal source, represented by effective temperature and chemical potential parameters. These effective thermodynamic properties are then extracted from the yields measured in the full phase space (the so-called $4\pi$-yields) \cite{Braun-Munzinger:1999hun,Begun:2018efg}. These $4\pi$-yields can be obtained by interpolating and extrapolating the measured yields within a few rapidity bins, and then integrating them over the rapidity variable $y$. We refer to this approach as the ``single source model.''

To derive the effective parameters $(T, \mu, V)$, these parameters are adjusted so that the identified hadron yields calculated using the thermal model match the corresponding measured $4\pi$-yields. The thermal model full phase space yields can be obtained by integrating out the rapidity variable $y$ in Eqs.~(\ref{eq:p_yield},\ref{eq:pik_yield}). In this case, the net proton yield is given by
\begin{align}
 \label{eq:nf}
N_{p-\bar p} =  \frac{g_pVT}{2\pi^2}m_p^2 K_2(\beta m_p) 2\sinh {\beta \mu}\,,
\end{align}
and the yields of pions and kaons read
\begin{equation} 
\label{eq:nb}
N_i =  \frac{g_iVT}{2\pi^2} m^2_i\sum_{n=1}^\infty \frac{1}{n}  K_2(n\beta m_i)\,,\quad i=\pi^+, K^+
\end{equation}
where $K_2(x)$ is the Bessel-$K$ function of the second kind and $\beta\equiv1/T$.

{\it \bf Discrete source model.} Another commonly used approach in thermal models is to assume that particles measured at rapidity $y$ are emitted from a thermal source at the same rapidity $y_s=y$, disregarding thermal smearing effects. In this scenario, when the thermodynamic properties $(T, \mu, V)$ of a thermal source at an arbitrary rapidity $y_s$ are known, the emitted hadrons are assumed to have the same rapidity, and their yields are determined by Eqs.~(\ref{eq:nf}, \ref{eq:nb}). For example, when analyzing the yields of identified particles at mid-rapidity, it is assumed that the extracted thermodynamic parameters $(T,\mu,V)$ obtained from Eqs.~(\ref{eq:nf}, \ref{eq:nb}) represent the properties of the system at $y_s=0$. This procedure is then repeated for measurements in different rapidity windows, allowing for the determination of thermodynamic parameters along the beam direction \cite{Begun:2018efg,Vovchenko:2019pjl}. This scenario of thermal models, where the thermal sources at different rapidities are considered independent from each other, is referred to as the ``discrete source model''. This scenario is straightforward to implement, but it neglects the thermal smearing effect, which could be significant when the thermodynamic properties of the fireball vary considerably along the beam direction, particularly at low beam energies.

{\it \bf Continuous source model.} Finally, we consider a more physical scenario which account for thermal smearing effects properly. In this case, a longitudinally inhomogeneous system can be treated as a superposition of thermal sources with varying thermodynamic properties in $y_s$, and each of them emits particles according to Eqs.~(\ref{eq:p_yield},\ref{eq:pik_yield}) with a longitudinal boost. If we represent the rapidity-dependent hadron distributions given by Eqs.~(\ref{eq:p_yield}, \ref{eq:pik_yield}) for a static source at $y_s{\,=\,}0$ collectively as $K_h\bigl[y;\,(T,\mu,V)({y_s{\,=\,}0})\bigr]$,\footnote{%
    To clarify, $h$ represents hadron species, including $p, \pi^+, K^+$ considered here. $K_h\bigl[y;\,(T,\mu,V)({y_s{\,=\,}0})\bigr]$ is given by Eq.~\eqref{eq:p_yield} for protons and by Eq.~\eqref{eq:pik_yield} for pions and kaons, respectively.
}
then the particle distributions originating from a boosted source located at any arbitrary $y_s$ can be derived as $K_h\bigl[y-y_s;\,(T,\mu,V)({y_s})\bigr]$ by shifting the two distributions in Eqs.~(\ref{eq:p_yield},\ref{eq:pik_yield}) accordingly. Then the rapidity-dependent particle yields due to the entire fireball are given by \cite{Becattini:2007qr,Becattini:2007ci,Cleymans:2007jj}
\begin{equation}\label{eq:had_rap}
    \frac{dN_h}{dy}\left(y\right) = \int_{{y_s}_\mathrm{min}}^{{y_s}_\mathrm{max}}  K_h\bigl[y-y_s;\,(T,\mu,V)(y_s)\bigr]\,dy_s\,,
\end{equation}
where the integral is carried out over the rapidity coverage of the system. The freeze-out parameters in the beam direction can be extracted by adjusting the thermodynamic profiles $(T,\mu,V)(y_s)$ in such a way that the rapidity-dependent particle yields, as given by Eq. \eqref{eq:had_rap}, match the experimental measurements. Since the particle yields at a given rapidity $y$ are emitted by thermal sources distributed continuously within a window of $y_s$, we refer to this approach as the ``continuous source model''.

We parametrize the profiles of the temperature and chemical potential using a Taylor expansion:
\begin{align}
    T(y_s)& = t_0 + t_2 y_s^2 + t_4 y_s^4 + \dots\,,\label{eq:para_T}\\
    \exp\left[\frac{\mu(y_s)}{T(y_s)}\right]&-\exp\left[-\frac{\mu(y_s)}{T(y_s)}\right] =z_0 + z_2y_s^2 + z_4 y_s^4 + \dots\,,\label{eq:para_mu}
\end{align}
where $t_0, t_2, \dots$ and $z_0, z_2, \dots$ are free parameters. We parametrize $\exp(\mu/T)-\exp(-\mu/T)$ in Eq.~\eqref{eq:para_mu} since it is roughly proportional to the net proton yields; more terms in the expansion are generally needed when $\mu$ varies more strongly along rapidity $y_s$. For the volume $V(y_s)$ we use a plateau with two half-Gaussian tails \cite{Shen:2020jwv,Du:2022yok}:
\begin{equation}\label{eq:para_V}
    V(y_s) = V_0\exp\left[-\frac{(|y_s|-y_c)^2}{2\sigma^2} \times\theta(|y_s|-y_c)\right]\,,
\end{equation}
where $V_0, y_c, \sigma$ are parameters that control the overall scale, plateau width, and tails' width. In contrast to previous studies~\cite{Becattini:2007qr,Becattini:2007ci,Cleymans:2007jj}, our approach does not impose the condition that $(T, \mu)$ lies on the chemical freeze-out line. This allows us to investigate whether the thermal model can accurately reproduce such a characteristic observed on the hydrodynamic freeze-out surface with the thermal yields sampled from it. It is worth noting that comparing the hydrodynamic freeze-out surface, defined in space-time coordinates, with the thermal model freeze-out profiles, which are parametrized in terms of rapidity, requires a mapping between $\eta_s$ and $y_s$. This mapping is dependent on the longitudinal flow, and we will delve into this aspect in Sec.~\ref{sec:compare}.

For a longitudinally homogeneous system, which produces flat rapidity distributions of identified particles, the three scenarios of the thermal model should give the same results for the thermodynamic parameters. As we shall show below, when the beam energy increases and the system becomes more homogeneous, the thermodynamic parameters extracted by the three scenarios are indeed more consistent (see Fig.~\ref{fig:thermal_pd} below).

%%%%%%%%%%%%%%%%%%%%%%%%%%%%%%%%%%%%%%%%%%%%%%%%%%%%%%%%%%%%%%%%%%%%%%%%%%%%%%%%%%%%%%%
\section{Results and discussion}\label{sec:results}
%%%%%%%%%%%%%%%%%%%%%%%%%%%%%%%%%%%%%%%%%%%%%%%%%%%%%%%%%%%%%%%%%%%%%%%%%%%%%%%%%%%%%%%

In this section, we first study the space-time distribution of thermodynamic quantities on the freeze-out surface within the hybrid model in Sec.~\ref{sec:hybrid}. Then we use the statistical thermal model to extract freeze-out profiles using the particle yields sampled at hadronization of the hybrid model in Sec.~\ref{sec:thermal}. The effects from the longitudinal flow and thermal smearing within the thermal model will be discussed in Sec.~\ref{sec:compare}.

%%%%%%%%%%%%%%%%%%%%%%%%%%%%%%%%%%%%%%%%%%%%%%%%%%%%%%%%%%%%%%%%%%%%%%%%%%%%%%%
\subsection{Freeze-out surface from hydrodynamics}\label{sec:hybrid}
%%%%%%%%%%%%%%%%%%%%%%%%%%%%%%%%%%%%%%%%%%%%%%%%%%%%%%%%%%%%%%%%%%%%%%%%%%%%%%%

\begin{figure}[!tbp]
\begin{center}
%\hspace{-.5cm}
\includegraphics[width=\linewidth]{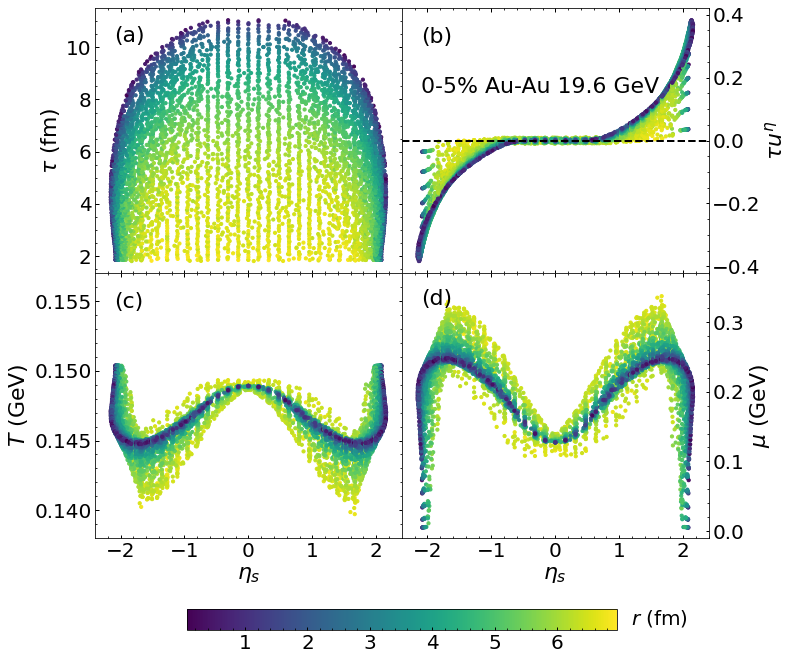}
    \caption{%
         Distributions in space-time rapidity \etas{} of freeze-out cells at $e_\mathrm{fo}=0.26$ GeV/fm$^3$ for 0-5\% Au+Au collisions at $\snn=19.6$ GeV: (a) longitudinal proper time $\tau$, (b) longitudinal flow $\tau{}u^\eta$, (c) temperature $T$ and (d) baryon chemical potential $\mu$. The colors represent the radial distance $r=\sqrt{x^2+y^2}$ of a fluid cell from the transverse center of the fireball. Cells with dark green are near the fireball center, and light yellow ones are towards the edge.}
    \label{fig:fzs_19}
\end{center}
\end{figure}

Using the constrained hybrid model discussed in Sec.~\ref{sec:hybrid_setup}, we obtain the freeze-out surface defined by a constant energy density  $e_\mathrm{fo}=0.26$ GeV/fm$^3$ \cite{Shen:2020jwv}.\footnote{\label{fn:efo}%
    The freeze-out energy density $e_\mathrm{fo}=0.26$ GeV/fm$^3$ used here is below 0.4 GeV/fm$^3$ used in Ref.~\cite{Denicol:2018wdp} which corresponds to the chemical freeze-out line extracted by the STAR Collaboration using thermal models \cite{STAR:2017sal}. This makes the freeze-out temperature around 150 MeV at zero chemical potential, below 160 MeV obtained in Ref.~\cite{STAR:2017sal} (see Fig.~\ref{fig:phase_diag_BES} below).
}  
We first focus on the distributions in space-time rapidity $\eta_s\equiv1/2\ln{[(t+z)/(t-z)]}$ for 0-5\% Au+Au collisions at $\snn=19.6$ GeV, shown in Fig.~\ref{fig:fzs_19}. In the figure, each scatter point represents a freeze-out fluid cell with energy density $e_\mathrm{fo}$. The point's color represents the radial distance $r=\sqrt{x^2+y^2}$ of a fluid cell from the transverse center of the fireball: Cells with dark green are near the fireball center, and light yellow ones are towards the edge. 

The space-time distribution of freeze-out cells is shown in Fig.~\ref{fig:fzs_19}(a), which is like a hemisphere. It indicates that at a specific \etas{}, fluid cells near the center of the fireball (dark green points) freeze out at later times, as expected, since the fireball has higher temperatures towards its center and, thus, it takes longer times to cool down to the freeze-out temperature. Moving from mid-space-time rapidity to the forward regions, with a decreasing initial energy density, the system freezes out earlier. The freeze-out surface expands slightly in \etas{} in the early expanding stage when $\tau\lesssim5\,$fm because of the longitudinal pressure gradients \cite{Du:2021zqz}. 

Fig.~\ref{fig:fzs_19}(b) shows the longitudinal flow velocity times proper time, $\tau u^\eta$, which would be zero in a Bjorken boost-invariant system with no longitudinal gradient. The plot shows that $\tau u^\eta$ increases from zero around mid- to forward-space-time rapidities, driven by the increasing longitudinal pressure gradients due to the inhomogeneity of the fireball in the beam direction. This indicates that the boost-invariance can be strongly broken away from midrapidity at such a low beam energy. Nevertheless, there is a window within $|\eta_s|\lesssim0.6$ where the boost-invariance is still a good approximation, corresponding to the longitudinal plateau region in the initial energy density where the longitudinal pressure gradient is small.\footnote{%
    The initial flow ignored in this study could affect the conclusions made here (see, e.g., Ref.~\cite{Shen:2020jwv}).
} 
At a specific \etas{}, the fluid cells near the fireball center can obtain larger longitudinal flow velocities, since it takes a longer time for them to freeze out, and thus the flow has more time to build up. 

%%%%%%%%%%%%%%%%%%%%%%%% - fig - %%%%%%%%%%%%%%%%%%%%%%%%
%%%%%%%%%%%%%%%%%%%%%%%%%%%%%%%%%%%%%%%%%%%%%%%%%%%%%%%%%

Figs.~\ref{fig:fzs_19}(c,d) show the distributions of temperature $T$ and baryon chemical potential $\mu$, which have ``W''- and ``M''-shapes, respectively. However, it should be noted that $T$ varies with rapidity by only about 5\% of its midrapidity value, while $\mu$ can vary by as much as 100\%. The feature of their shapes is a consequence of defining freeze-out surface at a constant energy density, for which $T$ and $\mu$ are anti-correlated through the EoS (see Fig.~\ref{fig:phase_diag_BES} below). The two humps in $\mu$ on the freeze-out surface originate from the ones in the initial baryon density, which generate the double-humped final net proton yields in Fig.~\ref{fig:bes_yields_tune}(b). This implies that extracting a $(T, \mu)$ from the $4\pi$-yields using the single source model would average out the variations in $T$ and $\mu$ along the beam direction; similarly for the case of measuring particle yields in a wide rapidity window. Additionally, even at a fixed \etas{}, $(T, \mu)$ of the freeze-out cells cannot be represented by a single point in the phase diagram, as $T$ can vary by about 5\% and $\mu$ even by about 30\%, at 19.6 GeV, because of the transverse inhomogeneity of the fireball. This indicates that, in the rapidity scan, the extracted $T$ and $\mu$ as a function of rapidity from rapidity-dependent particle yields are also effective values. However, compared to using the $4\pi$-yields, the rapidity scan can probe the phase diagram much more precisely (see Sec.~\ref{sec:compare} below).

%%%%%%%%%%%%%%%%%%%%%%%% - fig - %%%%%%%%%%%%%%%%%%%%%%%%
\begin{figure}[!tbp]
\begin{center}
%\hspace{-.5cm}
\includegraphics[width=1.01\linewidth]{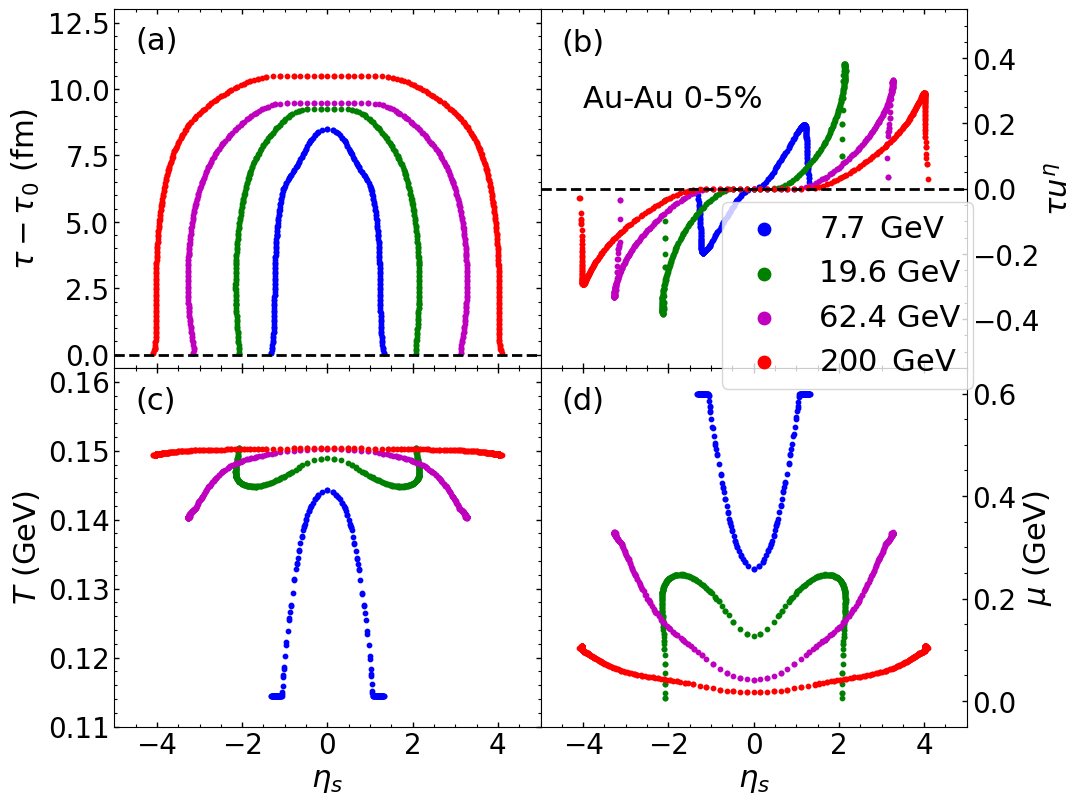}
    \caption{%
        Similar to Fig.~\ref{fig:fzs_19} but for 0-5\% Au+Au collisions at four beam energies: 7.7 (blue), 19.6 (green), 62.4 (magenta) and 200 (red) GeV. In panel (a), the initialization time $\tau_0$ has been subtracted from the proper time at different beam energies.
        }
    \label{fig:fzs_BES}
\end{center}
\end{figure}
%%%%%%%%%%%%%%%%%%%%%%%%%%%%%%%%%%%%%%%%%%%%%%%%%%%%%%%%%

Now we compare the distributions on the freeze-out surface for 0-5\% Au+Au collisions at the four beam energies in Fig.~\ref{fig:fzs_BES}. To illustrate the figure more clearly, we focus on the fluid cells around the center of the fireball, i.e., in the case of $\snn=19.6$ GeV, those darker green points in Fig.~\ref{fig:fzs_19}. Panel (a) shows the maximum time it takes for the system to freeze out along \etas{}. It indicates that, when the beam energy increases, the system takes a longer time to freeze out because of the higher initial temperature, and extends to larger \etas{} with larger beam rapidity. The central plateau region, where the boost-invariance is still a good approximation, gets wider when the beam energy increases. This is also clearly shown in panel (b), which compares the longitudinal flow $\tau u^\eta$: When the beam energy decreases, the window where $\tau u^\eta$ is zero shrinks and the maximum $\tau u^\eta$ increases, indicating a more strongly broken boost-invariance. However, at $\snn=7.7$ GeV, the maximum $\tau u^\eta$ is the smallest, mainly because of the shorter lifetime of the fireball and thus less time for the longitudinal flow to build up. The longitudinal flow still lacks quantitative constraints, and a wider fireball can compensate for a smaller longitudinal flow in principle along rapidity without changing the observables in Fig.~\ref{fig:bes_yields_tune}. We shall illustrate this further in Sec.~\ref{sec:compare} below.

%%%%%%%%%%%%%%%%%%%%%%%% - fig - %%%%%%%%%%%%%%%%%%%%%%%%
\begin{figure}[!tbp]
\begin{center}
%\hspace{-.5cm}
\includegraphics[width=\linewidth]{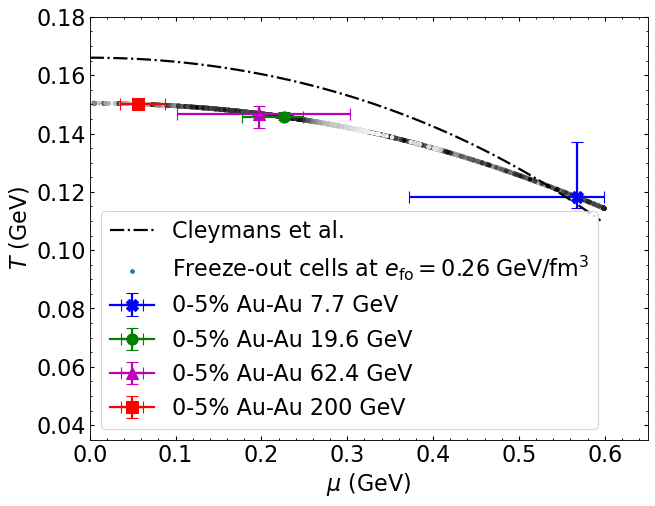}
    \caption{%
        Distributions of $(T, \mu)$ in the phase diagram for the fluid cells on the freeze-out surfaces from 0-5\% Au+Au collisions at the four beam energies. The grayish scatter points represent the freeze-out fluid cells at $e_\mathrm{fo}=0.26$ GeV/fm$^3$. The four markers with error bars illustrate the medians and 25\% and 75\% percentiles of $(T, \mu)$ distributions for the freeze-out cells at the four beam energies. As a comparison, the black dot-dashed line represents the chemical freeze-out line extracted using statistical thermal models from Ref.~\cite{Cleymans:2005xv}. }
    \label{fig:phase_diag_BES}
\end{center}
\end{figure}
%%%%%%%%%%%%%%%%%%%%%%%%%%%%%%%%%%%%%%%%%%%%%%%%%%%%%%%%%

Figs.~\ref{fig:fzs_BES}(c,d) show the $T(\eta_s)$ and $\mu(\eta_s)$ profiles on the freeze-out surface at the four beam energies, which share a general trend except at $\snn=19.6$ GeV. We see that $T(\eta_s)$ is the highest near mid-space-time rapidity and decreases toward forward- and backward-regions, and when the beam energy $\snn$ goes from 7.7, 62.4 to 200 GeV, $T(\eta_s)$ becomes flatter, wider and almost a constant at the top energy. In other words, the system becomes more isothermal and homogeneous when the beam energy increases. Correspondingly, because of the anti-correlation between $\mu$ and $T$ on the freeze-out surface, $\mu(\eta_s)$ is the lowest around midrapidity and rises when $|\eta_s|$ increases. The unique features at 19.6 GeV are related to the fact that the distance between the two humps in $dN^{p-\bar{p}}/dy$ is relatively smaller than the central plateau of $dN^\mathrm{ch}/d\eta$, compared to the other beam energies, as shown in Fig.~\ref{fig:bes_yields_tune}. Because of this, the distance between the two humps of the initial baryon density is smaller than the central plateau width in the initial energy density, which results in the ``W''- and ``M''-shapes for $T$ and $\mu$, respectively, on the freeze-out surface.  We have checked that manually increasing the distance between the two peaks in the initial baryon density at 19.6 GeV result in $T(\eta_s)$ and $\mu(\eta_s)$ on the freeze-out surface having similar shapes to those observed at the other energies.

From Figs.~\ref{fig:fzs_BES}(c,d), it is evident that $T(\eta_s)$ and $\mu(\eta_s)$ exhibit changes along space-time rapidity. Moreover, even at a fixed $\eta_s$, $T$ and $\mu$ can also spread out (Fig.~\ref{fig:fzs_19}) due to the inhomogeneity in the transverse direction. This is caused by the fact that the fireball is denser near the center and becomes more dilute towards the edge. To demonstrate the $(T, \mu)$ regions probed by the hadronization process at different beam energies, we plot the medians and 25\% and 75\% percentiles of $(T, \mu)$ distributions of all fluid cells on the freeze-out surface in the phase diagram shown in Fig.~\ref{fig:phase_diag_BES}. The markers in the figure represent the medians and error bars the first and third quartiles. The grayish scatter points represent $(T, \mu)$ of the fluid cells on the freeze-out surface at $e_\mathrm{fo}=0.26$ GeV/fm$^3$, which is below the freeze-out line extracted using statistical thermal models by Ref.~\cite{Cleymans:2005xv} corresponding to an energy density around 0.4 GeV/fm$^3$ (see footnote \ref{fn:efo} and also Ref.~\cite{Schafer:2021csj} using $e_\mathrm{fo}=0.5$ GeV/fm$^3$ which gives a phase transition line above that of Cleymans {\it et al}. \cite{Cleymans:2005xv}).

Fig.~\ref{fig:phase_diag_BES} shows that the $(T, \mu)$-medians lie on the freeze-out line, and when the beam energy decreases, the median of $T$ decreases and that of $\mu$ increases. This reminds us of how the $(T, \mu)$ values extracted by the STAR Collaboration using thermal models with midrapidity particle yields at BES energies lie on the chemical freeze-out line \cite{STAR:2017sal}. $(T, \mu)$ spread out following non-Gaussian distributions, indicated by the asymmetric error bars, which can contribute to the cumulants of final particle distributions. As shown by the figure, the fireball becomes less homogeneous at lower beam energies, and thus $(T, \mu)$ covers more extensive ranges. Hence using the single source model to extract $(T, \mu)$ from the $4\pi$-yields only gives effective thermodynamic properties of the fireball. We demonstrate this point further in the next section.

%%%%%%%%%%%%%%%%%%%%%%%%%%%%%%%%%%%%%%%%%%%%%%%%%%%%%%%%%%%%%%%%%%%%%%%%%%%%%%%
\subsection{Freeze-out profiles from thermal models}\label{sec:thermal}
%%%%%%%%%%%%%%%%%%%%%%%%%%%%%%%%%%%%%%%%%%%%%%%%%%%%%%%%%%%%%%%%%%%%%%%%%%%%%%%

In this section, we implement the thermal models discussed in Sec.~\ref{sec:thermal_setup} to extract the freeze-out profiles in rapidity using the identified particle yields given by the hybrid model in Sec.~\ref{sec:hybrid_setup}. As we have discussed, the thermal models assume that the hadrons are emitted from a thermally and chemically equilibrated source, and, in heavy-ion collisions, the extracted temperature and chemical potential using such models correspond to the properties at the chemical freeze-out. However, in heavy-ion collisions, resonances and heavier particles can decay after their thermal production, thus contributing to the final particle yields of lighter mesons and baryons measured experimentally. Thus, the implementation of thermal models must take into account the feed-down effect from the decays of unstable particles to the measured final hadron yields.\footnote{%
    Depending on the reconstruction efficiencies for particles in different experiments, decays from various interactions need to be considered accordingly. For example, the yields of $p$ and $\bar p$ measured in Ref.~\cite{STAR:2017sal} are inclusive, and thus the feed-down from weak decays needs to be considered for comparison.
} 

In practice, one way to implement thermal models is first starting with certain $(T, \mu, V)$, generating primordial hadrons whose multiplicities are given by hadron resonance gas models, then sampling their momenta according to, e.g., blast wave models, and finally letting the resonances decay through all decay chains until only the stable hadrons remain (see, e.g., Ref.~\cite{Vovchenko:2019pjl}). The yields of the final hadrons can then be compared to the measurements. Iterations are needed until the calculated final yields fit the experimental data, and the extracted $(T, \mu, V)$ should describe the thermodynamic properties at the chemical freeze-out. Microscopic transport models, which include the hadronic rescatterings, can also be used for such a purpose; see, e.g., Ref~\cite{Becattini:2012xb} which uses \urqmd. 

%%%%%%%%%%%%%%%%%%%%%%%% - fig - %%%%%%%%%%%%%%%%%%%%%%%%
\begin{figure}[!tbp]
\begin{center}
%\hspace{-.5cm}
\includegraphics[width=0.78\linewidth]{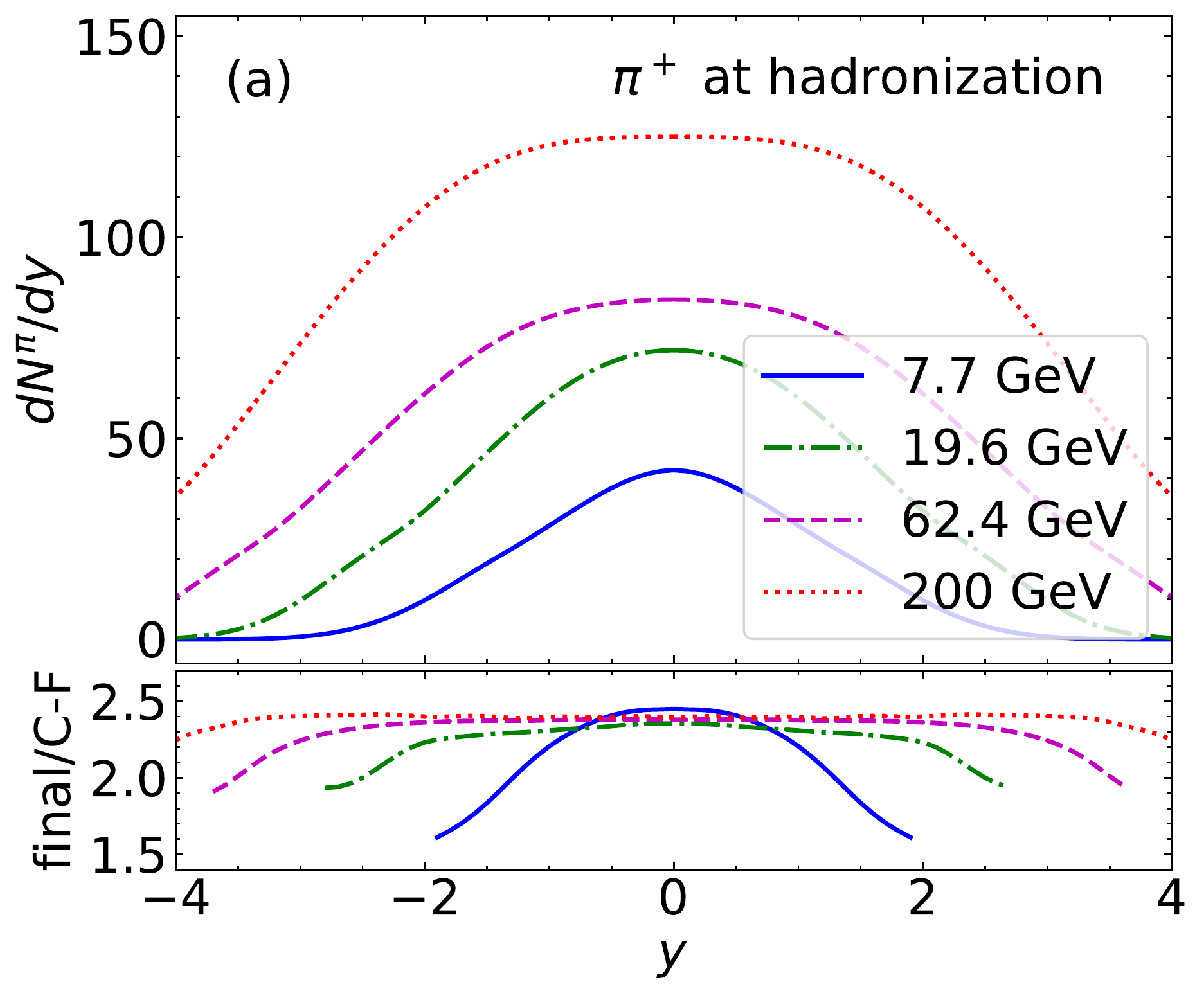}
\includegraphics[width=0.76\linewidth]{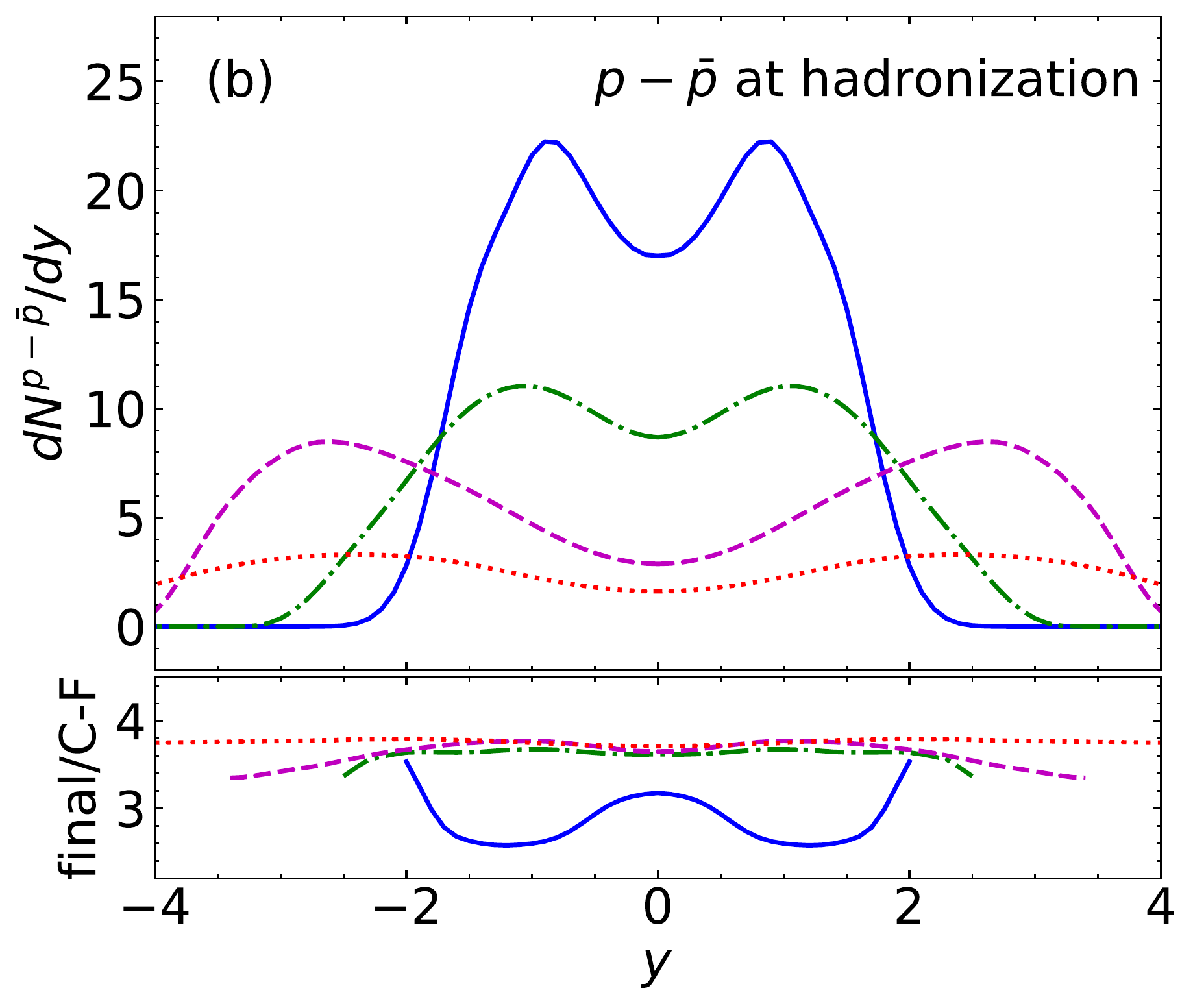}
\caption{%
    The yield change of identified particles during the hadronic stage for 0-5\% Au+Au collisions at 7.7 (blue solid), 19.6 (green dot-dashed), 62.4 (magenta dashed), and 200 (red dotted) GeV. The upper panels show the rapidity-dependent yields at hadronization obtained from the Cooper-Frye prescription for (a) pions and (b) net protons. The lower panels show the ratio between the final yields after the hadronic stage (``final'') and the Cooper-Frye yields in the upper panels at hadronization (``C-F'').}
    \label{fig:final_thermal}
\end{center}
\end{figure}
%%%%%%%%%%%%%%%%%%%%%%%%%%%%%%%%%%%%%%%%%%%%%%%%%%%%%%%%%

In this study, we take advantage of the hybrid model in Sec.~\ref{sec:hybrid_setup}, which can give the primordial hadron yields before decays with full coverage in rapidity at hadronization, shown in the upper panels of Fig.~\ref{fig:final_thermal}. In the lower panels, we also present the ratios between the final yields after the decays (``final'') and the yields sampled at the freeze-out surface using the Cooper-Frye prescription (``C-F''). For the two identified species, the ratios are pretty constant within a rapidity window at 19.6, 62.4, and 200 GeV, while they change significantly along rapidity at 7.7 GeV; the case of $K^+$ (not shown) is similar to that of $\pi^+$, with a ratio around 1.6 near midrapidity. This indicates that a constant conversion factor should not be used to estimate the thermal yields from the final ones when applying the rapidity scan for the low beam energies at $\snn\lesssim 10$ GeV. This also implies that the estimation of net baryon number from the net proton yields is highly rapidity-dependent at those low beam energies \cite{NA49:1998gaz,BRAHMS:2009wlg}.

%%%%%%%%%%%%%%%%%%%%%%%% - fig - %%%%%%%%%%%%%%%%%%%%%%%%
\begin{figure*}[!htbp]
\begin{center}
\hspace{-.5cm}
\includegraphics[width = \textwidth]{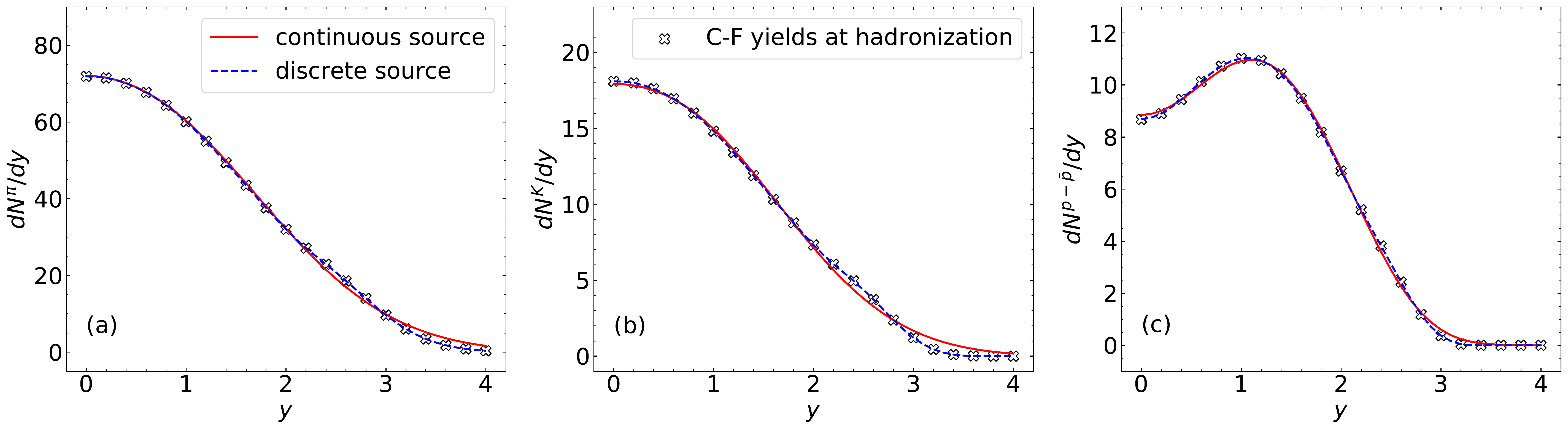}
\caption{Rapidity distribution of identified particle yields from two scenarios of the thermal model: (a) pions $\pi^+$, (b) kaons $K^+$ and (c) net protons $p-\bar p$, for 0-5\% Au+Au collisions at $\snn= 19.6{\rm \ GeV}$. The red solid line is for the continuous source model with thermal smearing, and the blue dot-dashed line is for the discrete source model without thermal smearing.  The open markers show the Cooper-Frye yields sampled at hadronization shown in Fig.~\ref{fig:final_thermal}.
}
\label{fig:19p6yields}
\end{center}
\end{figure*}
%%%%%%%%%%%%%%%%%%%%%%%%%%%%%%%%%%%%%%%%%%%%%%%%%%%%%%%%%

%%%%%%%%%%%%%%%%%%%%%%%% - fig - %%%%%%%%%%%%%%%%%%%%%%%%
\begin{figure*}[!htbp]
\begin{center}
\hspace{-.75cm}
\includegraphics[width = 1.018\textwidth]{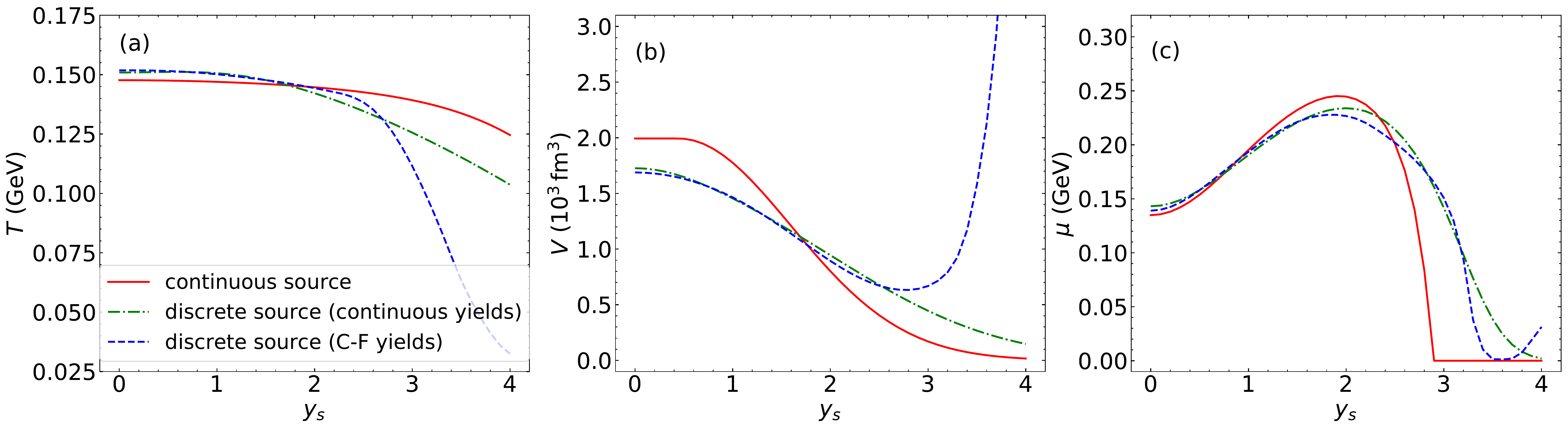}
\caption{Freeze-out profiles of (a) temperature $T$, (b) volume $V$, and (c) baryon chemical potential $\mu$ as functions of rapidity $y_s$ extracted by the two thermal models using the identified particle yields shown in Fig.~\ref{fig:19p6yields}. The green dot-dashed line corresponds to the profiles extracted by the discrete source model using the particle yields given by the continuous source model, i.e., the red solid lines in Fig.~\ref{fig:19p6yields}.}
\label{fig:19p6td}
\end{center}
\end{figure*}
%%%%%%%%%%%%%%%%%%%%%%%%%%%%%%%%%%%%%%%%%%%%%%%%%%%%%%%%%

Using the Cooper-Frye yields, we apply the thermal models described in Sec.~\ref{sec:thermal_setup} to extract the freeze-out profiles for 0-5\% Au+Au collisions at $\snn= 19.6{\rm \ GeV}$. Fig.~\ref{fig:19p6yields} shows the identified yields reproduced by the continuous (red solid line) and discrete (blue dashed line) source models, compared to the Cooper-Frye yields (black markers). As illustrated in Fig. \ref{fig:19p6yields}, the discrete source model gives an exact reproduction of the yields. This is because each rapidity is considered as a separate and independent thermal source, allowing for flexible adjustment of its thermodynamic properties. On the other hand, while the continuous model reproduces the Cooper-Frye yields around midrapidity quite well, minor discrepancies can be observed at forward-rapidities (at $y\gtrsim 2$). Complete removal of these discrepancies is not feasible due to the thermal smearing in the continuous model, which leads to correlations among the yields across different rapidities. Thus the adjustment of  thermodynamic profiles needs to consider the entire particle distributions along rapidity.

The thermodynamic profiles illustrated by the red solid and blue dashed lines in Fig.~\ref{fig:19p6td} correspond to the yields shown in Fig.~\ref{fig:19p6yields} using the same line styles. As the continuous model fails to completely reproduce the Cooper-Frye yields unlike the discrete model, we sought to enable a direct comparison between the two. Consequently, we applied the discrete source model to the yields obtained from the continuous model, resulting in the extraction of profiles indicated by the green dot-dashed lines in Fig.~\ref{fig:19p6td}. In the case of midrapidity ($y_s\lesssim 2$), the continuous source model exhibits a slightly smaller temperature, while the volume is approximately 15\% larger compared to the discrete source model. This observation can be attributed to the prefactor $VT^3$, which governs the overall scale in the rapidity distributions described by Eqs.~(\ref{eq:p_yield}, \ref{eq:pik_yield}). A small variation in temperature requires a significant adjustment in volume to fit the measured hadron yields. Overall, around midrapidity with $y_s\lesssim 2$, the two models give quite similar $(T, \mu)$. If uncertainties were introduced in the yields to represent experimental error bars, the difference between the two models would likely become unresolvable. 

At forward-rapidity, however, a much larger difference can be observed between the results of the two models. Compared to the discrete model, the continuous one gives a temperature decreasing more slowly towards forward-rapidities, implying a more isothermal system, and the volume decreases faster accordingly to give the decreasing yields. Similar observations can be made for the chemical potential. The net proton yield is nonzero at $y \gtrsim 3.5$ (red solid line in Fig.~\ref{fig:19p6yields}(c)) even when the chemical potential is zero at $y \simeq 2.95$ (red solid line in Fig.~\ref{fig:19p6td}(c)). This illustrates the thermal smearing effect included in the continuous model, accounting for the rapidity spread due to the thermal motion of the radiated particles.

Now we focus on the same discrete model fitted to slightly different yields from Fig.~\ref{fig:19p6yields}, represented by the green dot-dashed and blue dashed lines in Fig.~\ref{fig:19p6td}. This comparison is especially interesting, because it can be used to illustrate the model's sensitivity to the uncertainties in the yields that are unavoidable in experimental measurements. From Fig.~\ref{fig:19p6td}, we see that the extracted freeze-out profiles in the two cases are very similar when $y_s \lesssim 2.0$, where the two corresponding yields only have tiny deviations. However, they become quite different when $y_s \gtrsim 2.5$, where the two yields are slightly different. The blue dashed line in Fig.~\ref{fig:19p6td}(b) shows that the volume given by the discrete model can increase significantly when $y_s \gtrsim 3.0$, which is apparently unphysical.

The sensitivity of the temperature at forward-rapidities where the yields are small is an intrinsic property of the discrete model. This scenario obtains the temperature from the pion-to-kaon yields ratio $N_\pi/N_K$ using Eq.~\eqref{eq:nb}. Given the uncertainties of the two yields, $\delta N_\pi$ and $\delta N_K$, the extracted temperature would have the uncertainties given by
\begin{equation*}
\delta T = \frac{dT}{d (N_{\pi/K})} \delta \left(N_{\pi/K}\right)
= \frac{dT}{d (N_{\pi/K})} \frac{N_K \delta N_\pi - N_\pi \delta N_K}{(N_K)^2}\,,
\end{equation*}
where $N_{\pi/K}\equiv N_\pi/N_K$ denotes the ratio between the yields of $\pi^+$ and $K^+$.
A numerical estimation shows the factor $dT/d (N_{\pi/K})$ is about 0.01 GeV at forward rapidities. However, in the region at $y\gtrsim3$, the yield of kaons approaches 0, and thus small but nonvanishing $\delta N_\pi$ and $\delta N_K$ can cause a large $\delta T$, i.e., a significant change in the extracted temperature. $\delta N_\pi$ and $\delta N_K$ can be attributed to either the uncertainties in experimental measurements or imperfect fitting to the yields. In practice, because of the limited coverage of the detectors, the identified particle yields near the beam rapidities are not measurable. Thus significant theoretical uncertainties in the extracted freeze-out profiles would be unavoidable in these rapidity regions when the discrete model is implemented.

To briefly summarize, our results show that, near midrapidity (for $y_s\lesssim 2$ at 19.6 GeV), the continuous and discrete models extract relatively consistent freeze-out temperature and chemical potential. However, at forward-rapidities, the results can be significantly different, indicating large theoretical uncertainties in these regions.  In fact, when the yields are small, results from the discrete model can change dramatically and even become unphysical. On the other hand, the continuous model can avoid unphysical results by using smooth parameterized profiles that are adjusted to fit the entire rapidity data. Although the continuous model cannot reproduce the thermal yields exactly, accounting for the thermal smearing makes it a more physical model. In practice, the identified hadron yields are usually not measurable at forward or backward rapidities. However, the Event Plane Detector at STAR can measure charged hadrons in these regions \cite{Adams:2019fpo}, and this may help to reduce the theoretical uncertainties of the thermal models.

%%%%%%%%%%%%%%%%%%%%%%%% - fig - %%%%%%%%%%%%%%%%%%%%%%%%
\begin{figure}[!tbp]
\begin{center}
%\hspace{-.5cm}
\includegraphics[width=\linewidth]{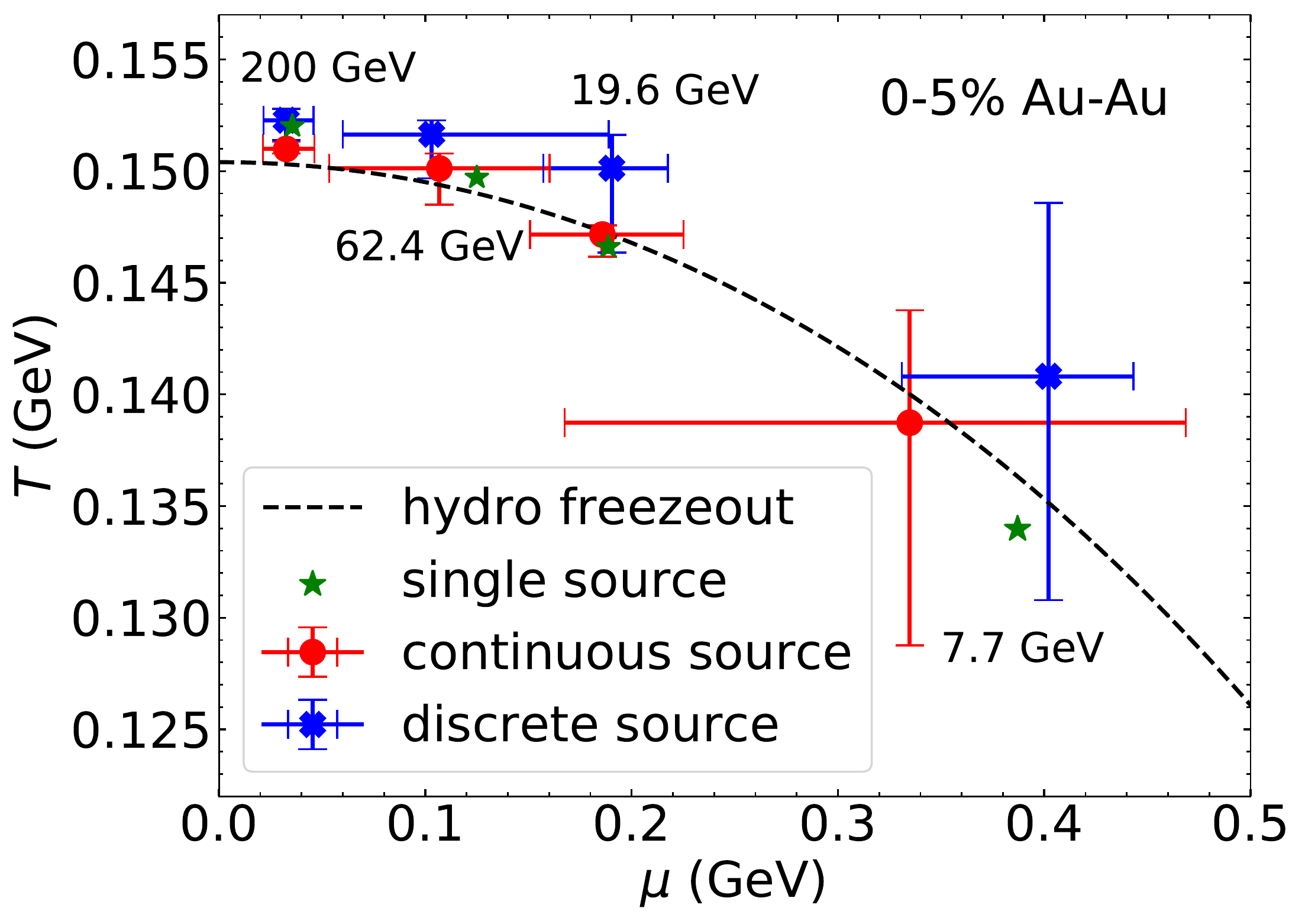}
\caption{Distributions of $(T, \mu)$ in the phase diagram extracted by different scenarios of the thermal models for 0-5\% Au+Au collisions at the four beam energies. The black dashed line shows a function $T(\mu)$ fitted to the hydrodynamic freeze-out line in Fig.~\ref{fig:phase_diag_BES}. The markers with error bars represent the median, and 25\% and 75\% percentiles of $(T, \mu)$ distributions for the two scenarios of the thermal model. The red circles represent the continuous source model, and the blue x-markers the discrete source model. The green stars also show the results obtained from the single source model for comparison.
}
\label{fig:thermal_pd}
\end{center}
\end{figure}
%%%%%%%%%%%%%%%%%%%%%%%%%%%%%%%%%%%%%%%%%%%%%%%%%%%%%%%%%

Finally, similarly to Fig.~\ref{fig:phase_diag_BES} for the hydrodynamic freeze-out surface, we show the distributions of $(T, \mu)$ in the phase diagram obtained by the thermal models for 0-5\% Au+Au collisions at four beam energies.  
To calculate the percentiles of the distributions, we take into account the contribution of the thermodynamic profile $\bigl(T(y_s), \mu(y_s)\bigr)$ at each $y_s$ by weighting it with the corresponding volume $V(y_s)$. In other words, the volume is treated as ``the number of fluid cells'' with $T(y_s)$ and $\mu(y_s)$ at rapidity $y_s$.
The results are shown in Fig.~\ref{fig:thermal_pd} for the continuous (red circle markers) and discrete (blue x-markers) models, which correspond to the red solid and blue dashed lines in Fig.~\ref{fig:19p6td} for the case of 19 GeV.\footnote{%
    The discrete model gives unphysically large rising volumes at forward-rapidities at $\snn = 7.7$ and 19.6 GeV, and these regions are excluded from the results. 
} 
The $(T, \mu)$ extracted by the single source model using the $4\pi$-yields are also shown  in the phase diagram, labeled by the green stars.

Fig.~\ref{fig:thermal_pd} shows that all three models give $(T, \mu)$ values around the hydrodynamic freeze-out line. The continuous model gives a slightly smaller temperature, and the median values of $(T, \mu)$ are more consistent with the hydrodynamic freeze-out line,  at all four beam energies, compared to the discrete model. At lower beam energies, the fireball becomes less isothermal and more inhomogeneous, which results in larger variations in the $\bigl(T(y_s), \mu(y_s)\bigr)$, as observed in the hydrodynamic model. This also explains why the $(T, \mu)$-point given by the single source model is further away from the median values given by the continuous model compared to the higher beam energies. Interestingly, although the single source model can only give effective $(T, \mu)$ of the entire fireball, they locate around the freeze-out line quite closely at all beam energies. This observation was also made by applying thermal models to the measured stable $4\pi$-yields at different beam energies \cite{Cleymans:2005xv}. Overall, the results given by the three models are more consistent when the fireball is more homogeneous at higher beam energies. Compared to the hydrodynamic results, the thermal models give a larger $T$ and a smaller $\mu$ at all beam energies by comparing Figs.~\ref{fig:phase_diag_BES} and \ref{fig:thermal_pd}.

%%%%%%%%%%%%%%%%%%%%%%%%%%%%%%%%%%%%%%%%%%%%%%%%%%%%%%%%%%%%%%%%%%%%%%%%%%%%%%%
\subsection{Longitudinal flow and thermal smearing}\label{sec:compare}
%%%%%%%%%%%%%%%%%%%%%%%%%%%%%%%%%%%%%%%%%%%%%%%%%%%%%%%%%%%%%%%%%%%%%%%%%%%%%%%

We have discussed the thermal smearing effect due to thermal motions of the radiated particles using the continuous source model. The hydrodynamic approach naturally implements such an effect in the Cooper-Frye prescription. However, it also includes the boost-non-invariant longitudinal flow discussed in Sec.~\ref{sec:hybrid}, which is not in the thermal model. This defers us from making a direct comparison for the freeze-out profiles of the two approaches.

Within the thermal models, the particle distributions in $y$ are determined by the thermodynamic profiles  in $y_s$ of the thermal source using Eq.~\eqref{eq:had_rap}. However, within the hydrodynamic approach, the thermodynamic distributions of fluid cells are in space-time rapidity $\eta_s$ as shown in Fig.~\ref{fig:fzs_19}. In other words, it implicitly involves another map between $\eta_s$ and $y_s$ to compare the two approaches. The map would be simply $y_s=\eta_s$, when the longitudinal flow is the boost-invariant Bjorken flow \cite{Bjorken:1982qr}. However, as shown in Figs.~\ref{fig:fzs_19} and \ref{fig:fzs_BES}, the pressure gradients drive longitudinal flows faster than the Bjorken flow, resulting in $y_s>\eta_s$, and the inequality is more substantial at larger $\eta_s$.

To make a direct comparison of the freeze-out profiles from the hybrid hydrodynamic model and the statistical thermal model, we shall introduce a parametrized longitudinal flow in the thermal models. Inspired by Fig.~\ref{fig:fzs_BES}(b), we parametrize the flow using a cubic function,
\begin{equation}\label{eq:uetas}
    \tau u^\eta(\eta_s)= \alpha(\eta_s-C)^3\,, 
\end{equation}
where $\alpha$ and $C$ are free parameters. Given such a flow profile, we can calculate the map between $y_s$ and $\eta_s$:
\begin{equation}\label{eq:yetas}
    y_s(\eta_s)=\frac{1}{2}\ln\frac{1+v^z(\eta_s)}{1-v^z(\eta_s)}\,,
\end{equation}
where $v^z=u^z/u^t$, and $u^t=u^\tau\cosh\eta_s+ (\tau u^\eta)\sinh\eta_s$ and $u^z=u^\tau\sinh\eta_s+(\tau u^\eta)\cosh\eta_s$; the normalization gives $u^\tau=\sqrt{1+(\tau u^\eta)^2}$ as the transverse flow is ignored in the thermal models. The longitudinal flow is incorporated in the thermal model by plugging Eq.~\eqref{eq:yetas} into Eq.~\eqref{eq:had_rap}, and the freeze-out parameters are functions of \etas{}, i.e., $(T, \mu, V)(\eta_s)\equiv(T, \mu, V)[y_s(\eta_s)]$. Using the same curve fitting methods, one can extract the freeze-out profiles in \etas{}.

%%%%%%%%%%%%%%%%%%%%%%%% - fig - %%%%%%%%%%%%%%%%%%%%%%%%
\begin{figure}[!btp]
\begin{center}
%\hspace{-.5cm}
\includegraphics[width=0.9\linewidth]{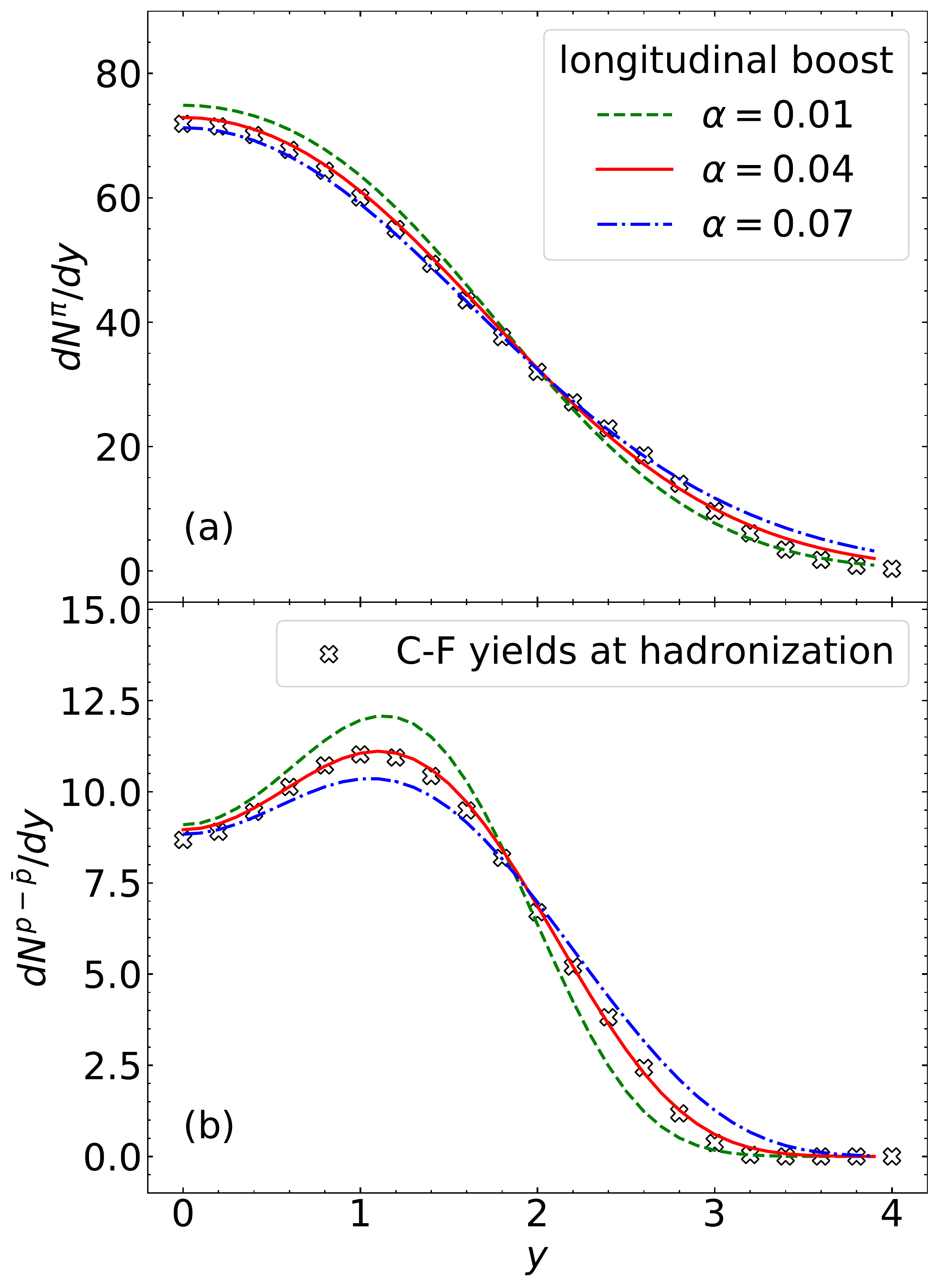}
\caption{Rapidity distributions of (a) pions and (b) net protons, for 0-5\% Au+Au collisions at $\snn=19.6$ GeV, obtained from the continuous source model with the same thermodynamic profiles $\bigl(\,T(\eta_s),\mu(\eta_s),V(\eta_s)\bigr)$ but under different boost-non-invariant longitudinal flows. The longitudinal flow is parametrized as $\tau u^\eta = \alpha(\eta_s-0.15)^3$ which fits the flow at $\snn=19.6$ GeV shown in Fig.~\ref{fig:fzs_BES}(b) when $\alpha=0.04$ (red solid). The values of $\alpha=0.01$ (green dashed) and $\alpha=0.07$ (blue dot-dashed) give smaller and larger longitudinal flows than $\alpha=0.04$, respectively. The black markers are the same Cooper-Frye yields shown in Fig.~\ref{fig:19p6yields}.
}
\label{fig:thermal_boost}
\end{center}
\end{figure}
%%%%%%%%%%%%%%%%%%%%%%%%%%%%%%%%%%%%%%%%%%%%%%%%%%%%%%%%%

At 19.6 GeV, Eqs.~(\ref{eq:uetas},\ref{eq:yetas}) with $\alpha=0.04$ and $C=0.15$ can give a flow profile which agrees very well with Fig.~\ref{fig:fzs_BES}(b). With such a longitudinal flow, we extract $(T, \mu, V)(\eta_s)$ profiles, and the corresponding particle distributions are shown by the red solid lines in Fig.~\ref{fig:thermal_boost}. To illustrate the effects of the longitudinal flow, the particle yield distributions are calculated for the cases of $\alpha=0.01$ (green dashed line, smaller flow) and $\alpha=0.07$ (blue dot-dashed line, larger flow) in Fig.~\ref{fig:thermal_boost}, with the same thermodynamic profiles in $\eta_s$. With a larger (smaller) longitudinal flow, the distributions get stretched (squeezed) in $y$, which is stronger for heavier species. This shows that for the same space-time rapidity distributions of $(T, \mu, V)$, the particle yields in rapidity are affected by the longitudinal flow.

\begin{figure}[!tbp]
\begin{center}
%\hspace{-.5cm}
\includegraphics[width=\linewidth]{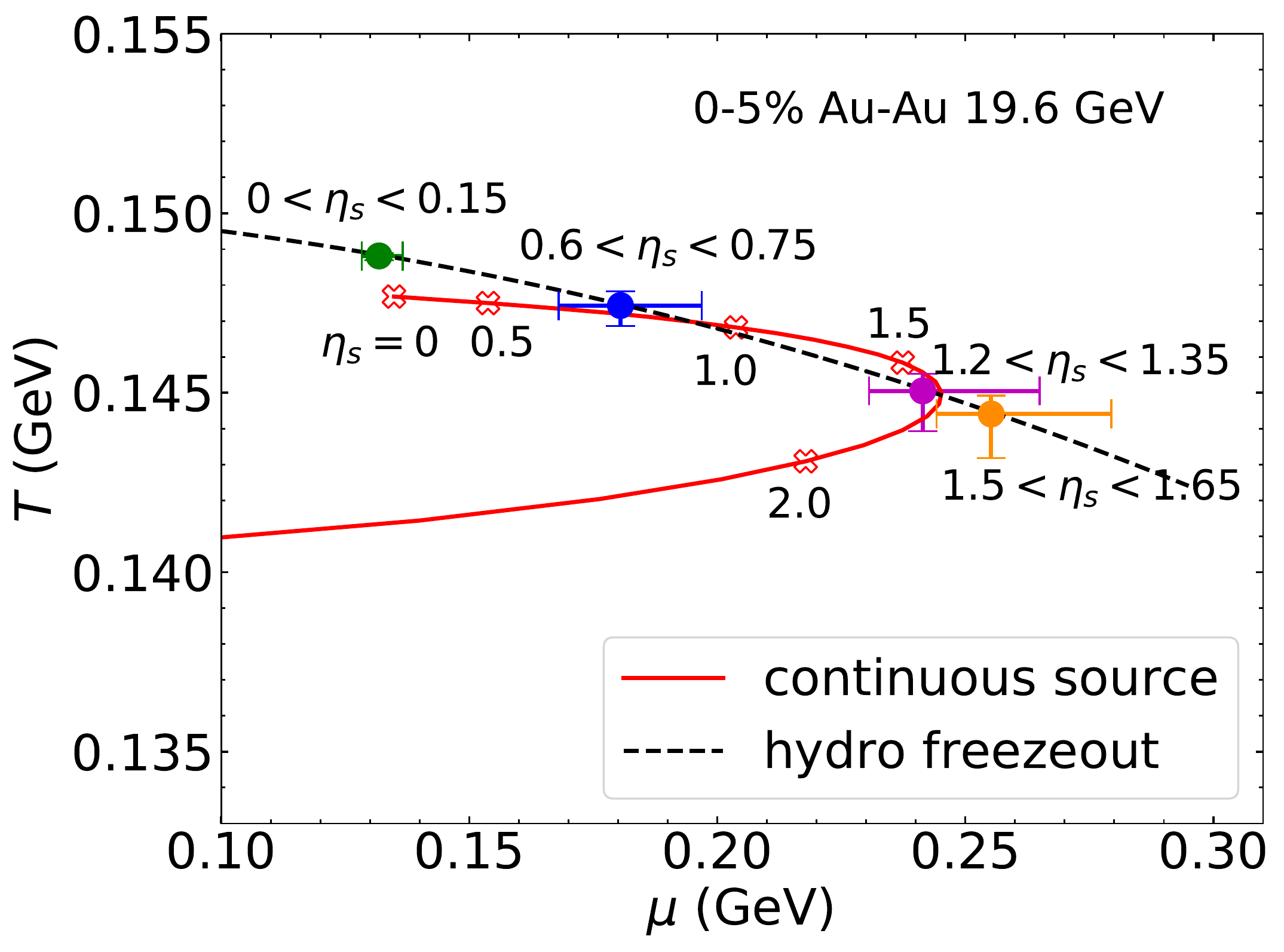}
\caption{Distributions of $(T, \mu)$ at different space-time rapidities in the phase diagram, obtained from the hydrodynamic freeze-out surface and the continuous source model, for 0-5\% Au+Au collisions at $\snn=19.6$ GeV. The black dashed line is the function $T(\mu)$ fitted to the hydrodynamic freeze-out line in Fig.~\ref{fig:phase_diag_BES}. The red solid line is obtained from the $(T, \mu)$ profiles in $y_s$ given by the continuous source model shown in Fig.~\ref{fig:19p6td}, and the red open x-markers show a few $\eta_s$ on the curve using the map $y_s(\eta_s)$ with $\alpha=0.04$. The circle markers with error bars are obtained in the same way as in Fig.~\ref{fig:phase_diag_BES}, but for freeze-out fluid cells within a few $\eta_s$ bins. 
}
\label{fig:pd_19p6_boost}
\end{center}
\end{figure}

Since $(T, \mu, V)(\eta_s)$ from both the hydrodynamic approach and the continuous source model are obtained, the thermodynamic properties along the beam direction can be compared directly. We make such a 
comparison in the phase diagram in the spirit of rapidity scan. Fig.~\ref{fig:pd_19p6_boost} illustrates the $(T, \mu)$ curve of the continuous model (red solid line) for 0-5\% Au+Au collisions at $\snn=19.6$ GeV, with a few x-markers indicating $\eta_s$ values. Fig.~\ref{fig:19p6td} shows that, from midrapidity towards larger $y_s$, the chemical potential first increases and then decreases at $y_s\simeq2$. Correspondingly, in Fig.~\ref{fig:pd_19p6_boost}, the curve is initially along the hydrodynamic freeze-out line towards higher $\mu$, then turns around to smaller $\mu$ at $\eta_s\simeq1.7$. However, the thermal model within $\eta_s\lesssim1.7$ gives a temperature that varies less strongly than on the hydrodynamic freeze-out curve. 

Within the hydrodynamic approach, the markers with error bars in Fig.~\ref{fig:pd_19p6_boost} indicate the medians and the first and third quartiles of $(T, \mu)$ for the freeze-out fluid cells within a few $\eta_s$ bins. Similar to the markers for the entire fireball at various beam energies in Fig.~\ref{fig:phase_diag_BES}, those for different $\eta_s$ bins of the fireball at the same beam energy also lie on the hydrodynamic freeze-out line, which is defined by a constant energy density. It is worth noting that the error bar in $\mu$ is much larger than that in $T$ due to the stronger variation of $\mu$ at a specific $\eta_s$ (see Fig.~\ref{fig:fzs_19}). As expected, the error bars are significantly smaller than those for the entire fireball in Fig.~\ref{fig:phase_diag_BES}, indicating that the rapidity scan can more precisely probe the phase diagram.

%%%%%%%%%%%%%%%%%%%%%%%% - fig - %%%%%%%%%%%%%%%%%%%%%%%%
%%%%%%%%%%%%%%%%%%%%%%%%%%%%%%%%%%%%%%%%%%%%%%%%%%%%%%%%%

While both the thermal model and the hydrodynamic approach give $(T, \mu)$ at different $\eta_s$ along the freeze-out line (for $\eta_s\lesssim1.7$), the same $(T, \mu)$ can correspond to different $\eta_s$ in the two models. Overall, the agreement of $(T, \mu)$ in $\eta_s$ is good, which indicates that the thermal model can successfully estimate the freeze-out surface for $\eta_s\lesssim1.7$ from the thermal yields sampled on it. This suggests that the thermal model incorporating thermal smearing and longitudinal flow performs well in the region close to midrapidity. However, there is a notable difference between the two models: The $(T, \mu)$ curve obtained from the thermal model generally shifts towards the $T$-axis at forward rapidities ($\eta_s\gtrsim1.7$) as both $T$ and $\mu$ decrease when the particle yields approach zero in these regions.\footnote{%
    In some early studies \cite{Becattini:2007qr,Becattini:2007ci}, the volume $V(y_s)$ is tuned to fit the rapidity distribution of $\pi^+$, and $\mu(y_s)$ to that of $p-\bar p$. $T(y_s)$ is obtained in such a way that the $(T, \mu)$ curve follows the chemical freeze-out line exactly. 
}
From the equation of state, we know that the region below the freeze-out line (dashed line in Fig.~\ref{fig:pd_19p6_boost}) has energy density below the freeze-out density $e_\mathrm{fo}$. Fig.~\ref{fig:pd_19p6_boost} indicates that the thermal model assumes the forward and backward rapidity regions (red solid curve below the turning point) to be in thermal equilibrium, even though the energy density is below the freeze-out density.

%%%%%%%%%%%%%%%%%%%%%%%% - fig - %%%%%%%%%%%%%%%%%%%%%%%%
\begin{figure}[!tbp]
\begin{center}
%\hspace{-.5cm}
\includegraphics[width=0.9\linewidth]{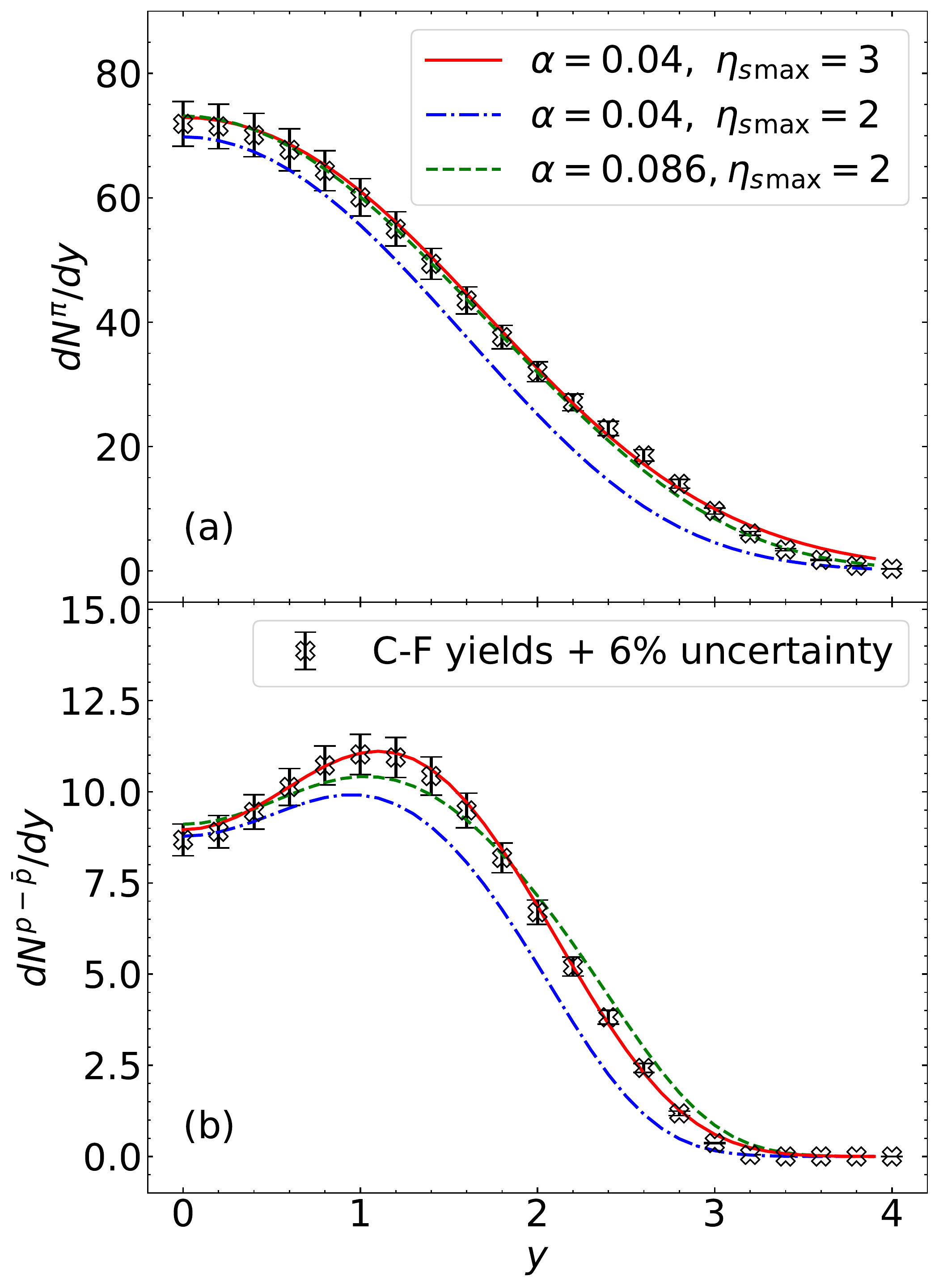}
\caption{Similar to Fig.~\ref{fig:thermal_boost}, with a 6\% uncertainty added to the Cooper-Frye yields represented by the error bars. The red solid line is the same as that in Fig.~\ref{fig:thermal_boost}, with the profiles of $(T, \mu, V)$ extended to ${\eta_s}_\mathrm{max}=3.0$. The blue dot-dashed and green dashed lines correspond to the same profiles of $(T, \mu, V)$ as the red line, which are cut at ${\eta_s}_\mathrm{max}=2.0$. Besides, the green line has a larger flow with $\alpha=0.086$ and a slightly larger volume scaled up by a constant factor 1.038.
}
\label{fig:thermal_boost_cut}
\end{center}
\end{figure}
%%%%%%%%%%%%%%%%%%%%%%%%%%%%%%%%%%%%%%%%%%%%%%%%%%%%%%%%%

This reminds us of another essential difference between the hydrodynamic and thermal models that has not been considered -- the longitudinal system size. Taking 19.6 GeV as an example, the system extends to ${\eta_s}_\mathrm{max}\approx3.0$ which corresponds to ${y_s}_\mathrm{max}\approx4.0$ (see Fig.~\ref{fig:19p6td}) in the thermal model. These profiles are obtained from the parametrizations in Eqs.~(\ref{eq:para_T}-\ref{eq:para_V}) with decreasing tails. On the other hand, the system described by hydrodynamics is distributed within $|\eta_s|\lesssim{\eta_s}_\mathrm{max}\approx2$ (see Fig.~\ref{fig:fzs_19}), one unit smaller than in the thermal model. 

To show the effects of the system size, we study a system with the same $(T, \mu, V)$ profiles in $\eta_s$ as the red solid line in Fig.~\ref{fig:thermal_boost} but are cut at $\eta_s=2.0$. The blue dot-dashed line shows the yields produced by such a system in Fig.~\ref{fig:thermal_boost_cut}. We see that the particle yields are reduced as expected but can still extend to $y\approx 3.5$ because of the longitudinal boost and thermal smearing effects. Now, if we increase the longitudinal boost by increasing $\alpha$ from 0.04 to 0.086 and scale up the volume $V(\eta_s)$ slightly by a constant factor of 1.038, we can see that the reproduced yields can still agree with the thermal yields with a 6\% uncertainty. This illustrates how a smaller system size in $\eta_s$ can be compensated by a more considerable longitudinal boost (with a larger volume). They cannot be disentangled easily using the rapidity-dependent yields with significant experimental uncertainties. Thus it is essential to consider the effects of thermal smearing, longitudinal flow, and system size when applying the thermal models to extract freeze-out parameters in the beam direction. Considering them would require more parameters in the thermal model, and its implementation is beyond the scope of this study. 

Some early studies also considered the effects from the ``longitudinal flow'' \cite{Schnedermann:1993ws,E895:2001zms,E-0895:2003oas,Netrakanti:2005iy}, which is, in fact, the Bjorken flow with $y_s=\eta_s$. These studies change ${\eta_s}_\mathrm{max}$, which effectively tunes the system size in space-time rapidity to change the width of rapidity distributions of particle species. This is similar to what is done here, but the longitudinal boost which results in $y_s>\eta_s$ was absent in those studies.

%%%%%%%%%%%%%%%%%%%%%%%%%%%%%%%%%%%%%%%%%%%%%%%%%%%%%%%%%%%%%%%%%%%%%%%%%%%%%%%
%%%%%%%%%%%%%%%%%%%%%%%%%%%%%%%%%%%%%%%%%%%%%%%%%%%%%%%%%%%%%%%%%%%%%%%%%%%%%%%%%%%%%%%
\section{Summary and conclusions}\label{sec:conclusions}
%%%%%%%%%%%%%%%%%%%%%%%%%%%%%%%%%%%%%%%%%%%%%%%%%%%%%%%%%%%%%%%%%%%%%%%%%%%%%%%%%%%%%%%

This study explored the rapidity scan method for heavy-ion collisions at BES energies within a multistage hydrodynamic framework and a statistical thermal approach. We calibrated the longitudinal bulk dynamics simulated by a (3+1)-dimensional multistage framework consisting of \music+\isd+\urqmd{}, using the pseudo-rapidity distribution of charged particles and net proton distribution in rapidity at four beam energies which cover an extensive range of the BES program. Such a framework was used to study the space-time distribution of thermodynamic properties of the nuclear matter on the freeze-out surface at the hadronization.

When the beam energy decreases, the system becomes more inhomogeneous in the beam direction, and the temperature and chemical potential of the fireball vary more strongly in the longitudinal direction, with a more significant variation for the latter. Even at a specific space-time rapidity, the fireball has finite variations in the thermodynamic properties because of its transverse inhomogeneity. Thus, at hadronization, the collision system probes finite regions of the phase diagram, even though the $(T, \mu)$ of each freeze-out fluid cell locate on a thin freeze-out line corresponding to a constant energy density. The longitudinal variations in thermodynamics indicate the necessity of the rapidity scan method, which also suggests that measuring observables in a wider rapidity window may increase the statistics but also blurs the probing of the QCD phase diagram.

Starting with the Bjorken flow, the pressure gradients due to the longitudinal inhomogeneity can induce longitudinal flows, which strongly break the boost invariant approximation. At the same time, a boost-invariant window still exists, which shrinks when the beam energy decreases. At forward rapidities, the system obtains flows faster than the Bjorken flow, and thus the rapidity of a fluid cell $y_s$ is greater than its space-time rapidity $\eta_s$. Because of the longitudinal boost from the flow and the rapidity spread due to thermal motions, particles can reach rapidities further than where the matter distributes in space-time rapidities. These two effects, together with the system size in $\eta_s$, are essential when explaining the particle yields at forward rapidities. 

Three thermal model scenarios are implemented to extract the freeze-out parameters from the hydrodynamic thermal yields. This allows for evaluating the performance of these thermal scenarios by comparing the extracted parameters to the hydrodynamic freeze-out profile. The goal is to explore the effectiveness of statistical thermal models in extracting freeze-out parameters using rapidity data and to identify some of their potential limitations. The continuous source model, which incorporates the dynamical features observed in the hydrodynamic model, exhibits more physical behavior at forward rapidities and demonstrates improved agreement with hydrodynamic results.

Near midrapidity, both the continuous source model, which considers the thermal smearing, and the discrete source model, which does not, give relatively consistent results for $(T, \mu)$. However, the results dramatically differ at forward rapidities, indicating significant theoretical uncertainties. In these regions where the particle yields become small, the discrete model becomes very sensitive to the yields and can give nonphysical results. The continuous model has a better behavior because the particle yields in rapidity correlate through thermal smearing.

The effect of the longitudinal flow is highlighted. Starting with the same $(T, \mu, V)$ profiles, a more significant longitudinal flow stretches the rapidity distributions more strongly. Thus, when applying the thermal model to extract the thermal parameters in rapidity, a more significant longitudinal flow can compensate for a smaller system size in $\eta_s$, and vice versa. More studies on thermal models are needed to explore how to disentangle the effects from the system size, thermal smearing, longitudinal flow, and $(T, \mu, V)$ distributions. 

The $(T, \mu)$ distributions for the freeze-out fluid cells from the hydrodynamic model and the extracted profiles using different scenarios of the thermal model are studied. For the case of the entire fireball at different beam energies, the $(T, \mu)$ medians of the former lie on the hydrodynamic freeze-out line. The large error bars corresponding to the 25\% and 75\% percentiles indicate the significant variations in $\mu$, originating from the longitudinal inhomogeneity in baryon density. The three scenarios of the thermal model also give medians of $(T, \mu)$ along the freeze-out line, while the continuous model agrees better with the freeze-out line and gives a slightly smaller temperature than the discrete model. However, we see that the median of $\mu$ is generally smaller, and that of $T$ is larger than what the hydrodynamic model gives at each beam energy. 

For the rapidity-binned system, the $(T, \mu)$ medians of the freeze-out cells within different rapidity bins also locate on the freeze-out line. However, the error bars become much smaller, indicating that the rapidity scan method can probe the phase diagram more precisely. On the other hand, although the curve of $(T, \mu)$ from the continuous source model follows along the freeze-out line in the regions near midrapidity, it starts to deviate from the freeze-out line and ends up on the $T$-axis at forward rapidities. However, cutting off the tails by reducing the system size in \etas{} together with the longitudinal flow and thermal smearing may help to remove this part of the curve and to improve its agreement with the hydrodynamic freeze-out line.

Finally, a thermal model that incorporates thermal smearing, system size, and longitudinal flow, inspired by the hydrodynamic approach, is proposed and investigated. By applying Bayesian analysis to the thermal models using upcoming BES-II data, one can constrain these effects without running expensive dynamical simulations. This can provide valuable guidance for analyzing data and constraining the dynamical modeling of nuclear collisions at low beam energies.

%%%%%%%%%%%%%%%%%%%%%%%%%%%%%%%%%%%%%%%%%%%%%%%%%%%%%%%%%%%%%%%%%%%%%%%%%%%%%%%%%%%%%%%
\section*{Acknowledgements}
%%%%%%%%%%%%%%%%%%%%%%%%%%%%%%%%%%%%%%%%%%%%%%%%%%%%%%%%%%%%%%%%%%%%%%%%%%%%%%%%%%%%%%%

The authors acknowledge fruitful discussions with Ulrich Heinz, Dmytro Oliinychenko, Chun Shen, Shuzhe Shi, and Volodymyr Vovchenko. This work was supported in part by the Natural Sciences and Engineering Research Council of Canada. Computations were made on the Beluga, Graham, and Narval computers managed by Calcul Quebec and by the Digital Research Alliance of Canada and the Ohio Supercomputer Center.

%%%%%%%%%%%%%%%%%%%%%%%%%%%%%%%%%%%%%%%%%%%%%%%%%%%%%%%%%%%%%%%%%%%%%%%%%%%%%%%%%%%%%%%
\bibliography{ref}
\end{document}